%% file: main.tex
\documentclass[a4paper, UKenglish, cleveref, numberwithinsect, thm-restate]{lipics-arxiv}

\hideLIPIcs
\pdfoutput=1

\usepackage[T1]{fontenc}
\usepackage{graphicx}
\usepackage{amsmath, amsfonts, amssymb}
\usepackage{mathtools}
\usepackage{multidef,xspace}
\usepackage{stmaryrd}
\usepackage{enumerate}
\usepackage{multirow}
\usepackage{cite}
\SetSymbolFont{stmry}{bold}{U}{stmry}{m}{n}
\usepackage{tikz}
\usetikzlibrary{automata, arrows.meta, positioning, arrows, calc, snakes}

\usepackage{hyperref}
\usepackage[capitalise]{cleveref}
\usepackage{color}

\usepackage{lmodern} %
\usepackage{enumitem}
\usepackage{todonotes}
\usepackage{mdframed}

\input{macros}

\title{Invariants for One-Counter Automata with Disequality Tests}

\titlerunning{Invariants for OCA with Disequality Tests}

\author{Dmitry Chistikov\footnote{%
During the work on this paper, DC was a visitor to the Max Planck Institute for Software Systems (MPI-SWS), Kaiserslautern and Saarbr\"ucken, Germany, a visiting fellow at St~Catherine's College and a visitor to the Department of Computer Science at the University of Oxford, United Kingdom.
}}
{Centre for Discrete Mathematics and its Applications (DIMAP) \&\newline{} Department of Computer Science, University of Warwick, Coventry, UK}
{d.chistikov@warwick.ac.uk} %
{https://orcid.org/0000-0001-9055-918X} %
{Supported in part by the Engineering and Physical Sciences Research Council [EP/X03027X/1].} %
 
\author{J\'{e}r\^{o}me Leroux}
{LaBRI, CNRS, Univ. Bordeaux, France} 
{jerome.leroux@labri.fr} %
{} %
{} %

\author{Henry Sinclair-Banks}
{Centre for Discrete Mathematics and its Applications (DIMAP) \&\newline{} Department of Computer Science, University of Warwick, Coventry, UK \and \url{http://henry.sinclair-banks.com}}
{h.sinclair-banks@warwick.ac.uk}
{https://orcid.org/0000-0003-1653-4069}
{Supported by EPSRC Standard Research Studentship (DTP), grant number EP/T51794X/1. Also supported in part by the International Emerging Actions grant (IEA'22) and by the ANR grant VeSyAM (ANR-22-CE48-0005).}

\author{Nicolas Waldburger}
{IRISA, Universit\'{e} de Rennes, France} 
{nicolas.waldburger@irisa.fr} %
{https://orcid.org/0009-0002-7664-5828} %
{Supported in part by the International Emerging Actions grant (IEA'22) and by the ANR grant VeSyAM (ANR-22-CE48-0005).} %

\authorrunning{D. Chistikov, J. Leroux, H. Sinclair-Banks, and N. Waldburger}

\Copyright{Dmitry Chistikov, J\'{e}r\^{o}me Leroux, Henry Sinclair-Banks, Nicolas Waldburger} 



\acknowledgements{
  We would like to thank Mahsa Shirmohammadi, without whom this work
  would not have been possible, for many ideas and encouragement.
  We are also grateful to Thomas Colcombet, Rayna Dimitrova, and James Worrell
  for useful discussions.
}

\nolinenumbers

\begin{document}

\maketitle

\begin{abstract}
\input{chapters/abstract.tex}
\end{abstract}

\section{Introduction}
\label{sec:introduction}
\input{chapters/introduction}

\section{Related Work}
\label{sec:related}
\input{chapters/related}

\section{OCA with Equality and Disequality Tests}
\label{sec:preliminaries}
\input{chapters/preliminaries}

\section{Getting Familiar with Disequality Tests}
\label{sec:bg-oca}
\input{chapters/background}

\section{Pessimistic Reachability}
\label{sec:pessimistic}
\input{chapters/pessimistic}

\section{Reachability in Strongly Connected OCA}
\label{sec:strongly-connected-OCA}
\input{chapters/strongly-connected}

\section{Combining Strongly Connected Components}
\label{sec:several-SCCs}
\input{chapters/leaky}

\section{Conclusions}
\label{sec:conclusions}
\input{chapters/conclusions}

\bibliography{references}

\begin{appendix}
\input{appendix}

\end{appendix}

\end{document}

%% file: macros.tex
\newtheorem{mynewdefinition}[theorem]{Definition}

\newtheorem{condition}[theorem]{Condition}

\newtheorem{numclaim}[theorem]{Claim}

\multidef{\textsf{#1}\xspace}{co,coNP,NP,PSPACE,coPSPACE,NPSPACE,PTIME,EXPTIME,EXPSPACE,NEXPTIME, PH}
\newcommand{\compclass}[1]{\textsf{#1}\xspace}
\newcommand{\Pcomplexity}{\textsf{P}\xspace}
\multidef[p=cal]{\ensuremath{\mathcal{#1}\xspace}}{A-Z}
\multidef[p=frak]{\ensuremath{\mathfrak{#1}\xspace}}{A-Z}

\def\NM@tuple@pen{\penalty-1000}
\def\NM@tuple@ldelim{\langle}
\def\NM@tuple@rdelim{\rangle}
\def\NM@tuple#1,#2,#3,#4\@NMendtuple{\def\NM@ux{#3}%
  #1
  \ifx\relax#2\relax\else,%
    \ifx\relax#3\relax\else
    \NM@tuple@pen@\let\NM@tuple@pen@\NM@tuple@pen
    \fi
    \NM@tuple #2,#3,#4\@NMendtuple
  \fi}
\def\tuple#1{\let\NM@tuple@pen@\relax
  \NM@tuple@ldelim\penalty10000\relax
  \NM@tuple #1,\relax,\relax,\relax\@NMendtuple
  \penalty10000\relax\NM@tuple@rdelim}

\newcommand{\newextmathcommand}[2]{%
    \newcommand{#1}{\ensuremath{#2}\xspace}
}

\def\set #1{\ensuremath{\{#1\}}}
\newcommand{\nats}{\mathbb{N}} 
\newcommand{\nset}[2]{[#1,#2]}

\newcommand{\VASS}{\mathcal{A}}
\newcommand{\rev}[1]{#1^{\textit{\textsf{R}}}}
\newcommand{\step}[1]{\xrightarrow{#1}}
\newcommand{\states}{Q}
\newcommand{\posstates}{Q_+}

\newcommand{\apath}{\sigma}
\newcommand{\transitions}{\MyDelta}
\newcommand{\atrans}{t}
\newcommand{\diseqtests}{\tau}
\newcommand{\guards}{\diseqtests}

\newcommand{\length}[1]{\mathsf{len}(#1)}
\newcommand{\effect}[1]{\mathsf{eff}(#1)}
\newcommand{\drop}[1]{\mathsf{drop}(#1)}

\newcommand{\poscycle}[1]{\gamma_{#1}}

\newcommand{\configs}{\Conf}

\newcommand{\posconfigs}{\configs_{\!+}}
\newcommand{\size}[1]{|#1|}
\newcommand{\norm}[1]{\|#1\|}
\newcommand{\achain}{\chain}
\newcommand{\run}{\rho}
\newcommand{\poststarpessimistic}[1]{\mathsf{Post}^*_-(#1)}
\newcommand{\postsymbol}{\mathsf{Post}}

\newcommand{\flowof}[1]{\mathsf{flow}(#1)}
\newcommand{\flowin}[2]{\mathsf{in}_{#1}(#2)}
\newcommand{\flowout}[2]{\mathsf{out}_{#1}(#2)}
\newcommand{\graphof}[1]{\mathcal{G}(#1)}
\newcommand{\aconfig}{c}
\newcommand{\expbound}{\psi}

\newcommand{\eqdef}{\overset{\mathrm{def}}{=}}
\newcommand{\defeq}{\eqdef}

\newcommand{\Conf}{\mathit{C\hspace{-2pt}o\hspace{-0.6pt}n\hspace{-1pt}f}}

\newcommand{\problemyy}[3]{
\par\noindent\underline{\textsc{#1}}\par\nobreak\vskip.2\baselineskip
\begingroup\clubpenalty10000\widowpenalty10000
\setbox0\hbox{Input:\ }\setbox1\hbox{Output:\ }
\dimen0=\wd0\ifnum\wd1>\dimen0\dimen0=\wd1\fi
\vskip-\parskip\noindent
\hbox to\dimen0{\box0\hfil}\hangindent\dimen0\hangafter1\ignorespaces#2\par
\vskip-\parskip\noindent
\hbox to\dimen0{\box1\hfil}\hangindent\dimen0\hangafter1\ignorespaces#3\par
\endgroup}

\newcommand{\Aa}{\mathcal{A}}
\newcommand{\Bb}{\mathcal{B}}
\newcommand{\Ii}{\mathcal{I}}
\newcommand{\Jj}{\mathcal{J}}

\newcommand{\NN}{\mathbb{N}}
\newcommand{\ZZ}{\mathbb{Z}}

\newcommand{\unstep}{\not\to}

\newcommand{\reach}[2]{#1\xrightarrow{*}#2}
\newcommand{\notreach}[2]{#1\: \rlap{\;\raisebox{0.33ex}{\tiny $/$}}{\xrightarrow{*}} \:#2}
\newcommand{\RUN}[3]{#1\xrightarrow{#2}#3}       
\newcommand{\candidate}[2]{#1\xrightarrow[\ZZ]{*}#2}

\newcommand{\notcandidate}[2]{#1\: \rlap{\;\raisebox{0.33ex}{\tiny $/$}}{\xrightarrow[\mathbb Z]{*}} \:#2}
\newcommand{\candidaterun}[3]{#1\xrightarrow[\ZZ]{#2}#3}

\newcommand{\Post}[1]{\mathsf{Post}(#1)}
\newcommand{\Pre}[1]{\mathsf{Pre}(#1)}

\newcommand{\pessimisticpoststar}[1]{\mathsf{Post}^*_-(#1)}
\newcommand{\pposts}[1]{\pessimisticpoststar{#1}}

\newcommand{\optimisticprestar}[1]{\mathsf{Pre}^*_+(#1)}
\newcommand{\opres}[1]{\optimisticprestar{#1}}

\newcommand{\bppostsymb}{\mathsf{LPost}^*_-}
\newcommand{\bpposts}[1]{\bppostsymb(#1)}

\newcommand{\bopres}[1]{\mathsf{LPre}^*_+(#1)}
\newcommand{\boprestar}[1]{\bopres{#1}}

\newcommand{\finv}{\forwardinvariant}
\newcommand{\binv}{\backwardinvariant}
\newcommand{\forwardinvariant}{\Ii}
\newcommand{\backwardinvariant}{\Jj}

\newcommand{\MyDelta}{\mathrm{\Delta}}
\newcommand{\MyPi}{\mathrm{\Pi}}
\newcommand{\MySigma}{\mathrm{\Sigma}}

\newcommand{\chain}{C}

\newcommand{\sset}{\subseteq}
\renewcommand{\emptyset}{\varnothing}
\newextmathcommand{\A}{\mathcal A}

\newcommand{\tempinit}{\source}

\newcommand{\source}{\mathsf{src}}
\newcommand{\target}{\mathsf{trg}}

\newcommand{\ind}{\textup{(\textsf{Ind})}\xspace}
\newextmathcommand{\pitwop}{\mathsf{coNP}^{\mathsf{NP}}}
\newextmathcommand{\sigmatwop}{\mathsf{NP}^{\mathsf{NP}}}
\newextmathcommand{\sigmathreep}{\MySigma_3^{\mathsf{P}}}
\newextmathcommand{\deltathreep}{\mathsf{P}^{\mathsf{NP}^{\mathsf{NP}}}}

\global\mdfdefinestyle{attractattention}{
linecolor = lipicsYellow, topline = false, bottomline= false, rightline = false, linewidth = 5pt
}

\newcommand{\posts}{\postsymbol}
\newcommand{\pres}{\mathsf{Pre}}

%% file: chapters/abstract.tex
We study the reachability problem for one-counter automata in which
transitions can carry disequality tests.
A disequality test is a guard that prohibits a specified counter value.
This reachability problem has been known to be \compclass{NP}-hard and in \compclass{PSPACE}, and
characterising its computational complexity has been left as a challenging
open question by Almagor, Cohen, Pérez, Shirmohammadi, and Worrell~\cite{AlmagorCPS020}.
We reduce the complexity gap, placing the problem into
the second level of the polynomial hierarchy, namely into the class $\pitwop$.
In the presence of both equality and disequality tests, our upper bound is
at the third level, $\deltathreep$.

To prove this result, we show that non-reachability can be witnessed by
a pair of invariants (forward and backward).
These invariants are almost inductive.
They aim to over-approximate only a ``core'' of the reachability set
instead of the entire set.
The invariants are also leaky: it is possible to escape the set. 
We complement this with separate checks as the leaks can only occur in a controlled way.

%% file: chapters/introduction.tex
It is well known that the computational complexity
of problems is often sensitive to seemingly minor adjustments in the problem setting.
Consider, for example, vector addition systems with states (VASS).
Perhaps more commonly presented as Petri nets, VASS are a very simple yet powerful model of concurrency.
Many important computational problems in logic, language theory, and formal verification
reduce to or are even equivalent to the reachability problem in VASS
(see, e.g.,~\cite{DBLP:journals/siglog/Schmitz16,hack75}).
However, a classical result due to Minsky shows that adding the capability to test counters for zero makes
the problem undecidable~\cite{minsky}.

At the same time, while reachability in VASS is known to be decidable~\cite{Mayr84},
its computational complexity was
recently shown to be extremely high, namely
Ackermann-complete (see~\cite{DBLP:conf/lics/LerouxS19} for the upper bound
and~\cite{DBLP:journals/jacm/CzerwinskiLLLM21,DBLP:conf/focs/Leroux21,DBLP:conf/focs/CzerwinskiO21} for the lower bound),
so from the practical point of view one might question the significance of the complexity jump arising from zero tests.

More recent, ``down-to-earth'', and perhaps more striking is the following result on 1-dimensional VASS,
which can be thought of as finite-state automata equipped with one counter (capable of storing a nonnegative integer).
Reachability in these systems can be decided in \NP~\cite{HaaseKOW09} and is in fact \NP-complete.
It is not difficult to show, using the standard hill-cutting technique~\cite{ValiantP75}, that
reachability can also always be witnessed by executions in which all values assumed by the counter
are bounded from above by an exponential function in the size of the system and the bit length of counter values
of the source and target configurations.
Because of this, one might expect that placing an exponential bound on the counter values
\emph{upfront} does not change the problem much.
But, in fact, the complexity jumps: the problem~--- which is equivalent to reachability in
two-clock timed automata~\cite{BO17}~--- becomes \PSPACE-complete~\cite{FearnleyJ15}.
One may say that, in this case, formal verification toolkit available for the reachability problem
is not robust to this change in the problem setting.

In this paper we study a different seemingly benign variation of the standard reachability question.
Consider one-counter automata in which transitions may test the value of the counter for \emph{disequality}
against a given integer (which depends on the transition). In other words, executions of the system can be blocked by
disequality guards, which \emph{prevent} the transition from being fired if the counter value is equal
to a specified number.
The initial motivation for studying this question comes from a model checking problem
for flat Freeze LTL;
see Demri and Sangnier~\cite{DemriS10} and Lechner, Mayr, Ouaknine, Pouly, and Worrell~\cite{LechnerMOPW18}.
Additionally, recall that automata can be used for the modeling of imperative code;
see, e.g., Hague and Lin~\cite{HagueL11,HagueL12}, as well as discussion in Section~\ref{sec:related}.
Classical Minsky machines encode \emph{if}--\emph{then} conditionals with equality comparisons to constants, $x = k$.
Simulating an \emph{if}--\emph{then}--\emph{else} conditional of this type on a Minsky machine seems to require
additional $O(k)$ states.
If $k$ is large, this growth in size may be exponential,
even though \emph{then} branches as well as increments $x\,\texttt{+=}\,k$ and decrements $x\,\texttt{-=}\,k$
can be encoded directly.
(If the machine model is extended with $x \leq k$ comparisons, then the asymmetry between
\emph{then} and \emph{else} disappears, but reachability becomes \PSPACE-complete~\cite{FearnleyJ15}.)

One might expect that, since only a small number of configurations are forbidden (by the disequality guards) in the infinitely large
configuration space, the complexity of the problem should not change significantly and existing techniques should be applicable.
This conclusion, however, has remained elusive.
For the problem of reachability in one-counter automata with disequality tests,
the exponential upper bound on counter values necessary for witnessing reachability carries over.
But, despite progress on related problems~\cite{BolligQS19,BlondinLMOP23}
(including the settling of the complexity of the above-mentioned flat Freeze LTL model checking problem~\cite{BolligQS19}),
it has not been possible to pin down the complexity of this problem,
which has been known to be \NP-hard (even without disequality tests) and to belong to \PSPACE (thanks to the
exponential bound on the counter)~\cite{LechnerMOPW18}.
The apparent simplicity of the problem contrasts with the lack of robustness
of the available toolbox.
It was recently shown by Almagor, Cohen, P\'erez, Shirmohammadi, and Worrell~\cite{AlmagorCPS020}
that the coverability (or state reachability) problem
can be solved in polynomial time for this model, similarly to the standard 1-dimensional VASS without tests.
The algorithm and its analysis, however, become sophisticated, and
the complexity of reachability was left as a \emph{challenging open problem}.

In the present paper, we make progress on this problem.
Existing techniques need to be extended and developed significantly to handle seemingly benign disequality tests.
We have been unable to find an easily verifiable witness for reachability, and instead show that
\emph{non-reachability} is witnessed by a form of invariants (or, more precisely, separators).
The existence of counterexamples that violate such invariants can be checked in \NP,
thus placing the reachability problem for OCA with disequality tests into the second level of the polynomial hierarchy, namely in $\pitwop$.
This complexity class captures the complement of synthesis-type questions, which ask to find
a single object (say, a circuit) that works correctly for all (exponentially many) inputs.
In our problem, one configuration can reach another if and only if every potential invariant
(of a form we describe) violates one of the invariance conditions; moreover, this violation can be checked in \NP.
In the presence of both equality and disequality tests, we need a slightly larger class $\deltathreep$,
at the third level of the hierarchy.

\subparagraph*{Our contributions.}
Traditionally, an invariant is an overapproximation of the set of reachable configurations
which is inductive, i.e., closed under the transition relation.
Our invariants are different in several ways:
\begin{enumerate}
	\item
		We capture only some configurations within the reachability set, which form its \emph{core}.
		Accordingly, we require a tailored notion of closure, namely closure under a restricted form of
		reachability relation.
	\item
		Our invariants are leaky (\emph{almost} inductive): an execution
		may escape the set.
		Allowing leaks is complemented by a separate controlling mechanism (check)
		that all leaks~--- which may occur at the interface between strongly connected components
		of the automaton~---
		are safe.
	\item
		To compose our local inductive invariants, i.e., those restricted
		to a single strongly connected component, the controlling mechanism
		for leaks relies on relaxed integer semantics for the execution.
		More precisely, we extend (to automata with disequality tests)
		a known technique~\cite{HaaseKOW09} for lifting
		$\mathbb Z$-executions to actual executions.
\end{enumerate}
Our notion of local invariants requires that we place
a certain technical assumption at the interface (entry and exit points) of strongly connected
components.
To discharge this assumption, we use a combination of
two invariants, one for the main (forward) VASS and another
for the reverse VASS.
Together, these two sets form a separator~--- a witness for non-reachability.

%% file: chapters/related.tex
\subparagraph*{Invariants.}
In formal verification, a forward exploration of countably infinite configuration spaces from the initial configuration, or a symmetrical backward exploration from the target, is a standard approach to reachability problems and targets bug finding. General heuristics can be used to improve such an exploration (see, for instance, the recent directed reachability algorithm~\cite{DBLP:conf/tacas/BlondinHO21}). However, in order to prove non-reachability, thus certifying the absence of bugs, an invariant-based approach is more popular.

Many techniques have been developed in the past for computing inductive invariants, depending on the structure of the underlying system based on counterexample-guided abstraction refinement~\cite{DBLP:conf/cav/ClarkeGJLV00}, automata~\cite{DBLP:conf/cav/HeizmannHP13}, property-directed reachability~\cite{DBLP:conf/vmcai/Bradley11,DBLP:conf/sat/Bradley12,AmatDH22}, and more generally in the abstract interpretation framework~\cite{DBLP:conf/popl/CousotC77}.

In vector addition systems, semilinear invariants~\cite{GS-PACIF66} are sufficient for the general reachability problem~\cite{Leroux10}. Even if those invariants are intractable in general, for some instances, namely the control-state reachability problem, the implementation of efficient tools computing invariants (downward-closed sets in that case) is an active research area~\cite{EsparzaLMMN14,BlondinFHH17} with implementation of tools~\cite{DBLP:conf/concur/KaiserKW12,DBLP:conf/tacas/BlondinFHH16,DBLP:conf/fossacs/FinkelHK20}.

In this paper we focus on 1-dimensional VASS in the presence of equality and disequality tests; we call them \emph{one-counter automata \textup{(OCA)} with tests}.
The notion of local inductive invariants with leaks, which we propose, provides a way to reduce the
search space of inductive invariants, by specifying the shape of the `core' of the invariant (a union of arithmetic progressions within `bounded chains'), as well as restricting the problem to each strongly connected component one by one. We view this as a compositional approach for computing inductive invariants. As a theoretical application, we prove that the reachability problem for one-counter automata with tests is between \NP and $\deltathreep$, and in fact in $\pitwop$ if only disequality tests are present.

The previous work on OCA with disequality tests by Almagor~et~al.~\cite{AlmagorCPS020} enables us to focus
(subject to a technical assumption)
on configurations in a small number
of bounded chains (see Section~\ref{sec:bg-oca}).
The structure of the set of reachable configurations in these chains admits a short description.
At the core of our invariants are exactly such sets, and we need an appropriate notion of ``inductiveness'',
a condition to control ``leaks'' that violate the assumption above, and a verification mechanism for all these conditions.

\subparagraph*{One-counter automata.}
OCA can be seen as an abstraction of pushdown automata, a widely used model of recursive systems.
Conceptually very simple, OCA are at the heart of a number of results in formal verification; see, e.g.,
\cite{LLT05,AlurC11,BaumannGMTZ23}.
Multi-counter automata are used to model imperative code with numerical data types~\cite{HagueL11,HagueL12};
roughly speaking, a reachability query is expressed in logic, as a formula in existential linear integer arithmetic. 
In these two references an additional pushdown stack is available, capturing recursive function calls.
We refer the reader to~\cite{ChistikovCHPW19} for a retrospective on
underlying `pumping' results for OCA, crucial for many of the recent results.
There is also a rich history of research on behavioural equivalences and model checking for
a variety of one-counter processes and systems; see, e.g.,~\cite{ValiantP75,BohmGJ13,GollerL13,StewartEY15}.

The above-mentioned result that reachability in OCA is NP-complete, by Haase, Kreutzer, Ouaknine, and Worrell~\cite{HaaseKOW09},
has recently been built upon to give a representation of the entire reachability relation
in existential linear integer arithmetic, with an implementation available online, by Li, Chen, Wu, and Xia~\cite{LiCWX20}.
The idea of `lifting' candidate runs to actual runs, which is shown in~\cite{HaaseKOW09}
and which we develop further by adding support for disequality tests,
has been used in other settings as well~\cite{DBLP:conf/fsttcs/Leroux19,DBLP:journals/jacm/BlondinEFGHLMT21,Mayr84}.
For example, a construction similar to our \Cref{unbounded_rev_unbounded_global}
is an element of the proof of a tight upper bound
on the length of shortest runs in OCA without disequality tests~\cite{ChistikovCHPW19}.
In comparison to the latter paper,
our construction need not consider divisibility properties of run lengths,
but at the same time applies in a more general scenario: the updates of our OCA are specified
in binary notation (that is, succinctly); and, naturally, our OCA may have disequality tests.

We already mentioned above that,
despite appearing atypical at first glance, the disequality tests do in fact contribute to the modeling power:
namely, when modeling code,
these tests enable the simulation of the \emph{else} branch in conditional statements comparing an integer variable for equality with some constant.
The framework of Hague and Lin~\cite{HagueL11,HagueL12}
assumes that each counter variable can undergo at most $k$ reversals
(i.e., changes between `increasing' and `decreasing'), where $k$ is fixed. This assumption is strong;
without it, a reachability instance would require a logical formula of exponential size.
Results of Haase~et~al.~\cite{HaaseKOW09} and Li~et~al.~\cite{LiCWX20} avoid this assumption,
but, for the standard syntax of one-counter automata, \emph{if}--\emph{then}--\emph{else} conditionals
remain out of reach --- or rather require an exponential expansion of the automaton.
Our leaky invariants technique allows us to handle such conditionals
with equality tests on counters, without assuming any bound on the number of reversals.

%% file: chapters/preliminaries.tex
We denote by $\ZZ$ and $\NN$ the set 
of all integers and all nonnegative integers, respectively.

A \emph{constraint} is either an \emph{equality test} of the form $x = k$ with $k \in \NN$, a \emph{disequality test} of the form $x \ne k$ with $k \in \NN$, or simply $\texttt{true}$; $x$ denotes here our counter, which is a nonnegative integer variable. Let $\mathcal{C}$ denote the set of all possible constraints.  
A \emph{one-counter automaton (OCA) with equality and disequality tests} is a triplet~$\VASS=(Q,\transitions,\tau)$, where $Q$ is a finite set of \emph{states}, $\transitions \subseteq Q\times \ZZ \times Q$ is a finite set of \emph{transitions} and $\tau \colon Q\to  \mathcal{C}$ is the \emph{constraint function}.
The automaton~$\VASS$ is an \emph{OCA  with  disequality tests}  if the constraint function $\tau$ does not have any equality tests. We sometimes refer to the constraints as \emph{guards}. 

Syntactically, $\VASS$ can be seen as an integer-weighted graph with directed edges between states. 
Viewed this way, $\VASS$ can be decomposed into a set of \emph{strongly connected components (SCCs)}. The automaton $\VASS$ is \emph{strongly connected} when it has one strongly connected component only.
A \emph{path} $\pi$ in $\VASS$ is a sequence $\pi = (t_1, t_2, \ldots, t_{n})$ of transitions, where $t_i = (q_{i-1}, a_i, q_{i})$ for each $i$ and $n \geq 0$.
We may refer to $\pi$ as a $q_0$--$q_{n}$ path. 
The \emph{length} of such a path is $\length{\pi} \eqdef n$.
The \emph{effect} of $\pi$ is $\effect{\pi}\eqdef \sum_{i=1}^n a_i$.
A \emph{cycle} is a path starting and ending at the same state; for $q \in Q$, a $q$-\emph{cycle} is a $q$--$q$ path.
A path or cycle is \emph{simple} if it contains no repetition of states, except that a \emph{simple cycle} has the same starting and ending state.
Every simple cycle has length less than or equal to $\size{Q}$.

The size of an OCA $\VASS$ is the bit size of its encoding, where all numbers are written in binary;
we denote it by $\size{\VASS}$.
We write $\norm{\transitions}$ and $\norm{\tau}$ to refer to the maximum absolute value of a transition update and test, respectively. 

\subparagraph*{Configurations and runs.}
The semantics of~$\VASS$ is defined based on the set of valid configurations and the reachability relation, as follows.

A \emph{configuration} is a pair $(q,z)$ comprising a state~$q\in Q$ and a nonnegative integer~$z\in \mathbb{N}$; 
we may refer to~$z$ as the \emph{counter value}. 
We say that~$(q,z)$  is a  \emph{valid configuration} if it respects the constraint $\tau(q)$.
Write $\Conf\eqdef Q\times \NN$ for the set of all configurations. Given two configurations $(q,z), (q',z')$ and $\atrans \in \transitions$, we write $(q,z) \step{\atrans} (q',z')$ when $\atrans = (q,z'-z,q')$; we denote by $(q,z) \step{} (q',z')$ the existence of such a transition. 
A \emph{run} of $\VASS$ is a sequence~$(q_0,z_0), \ldots, (q_n,z_n)$  of valid configurations, for $n \ge 0$, such that there exists a path $(t_1, \dots ,t_n)$ with $(q_{i-1},z_{i-1}) \step{t_i} (q_i,z_i)$.  %
We say that $(q_n,z_n)$ is \emph{reachable} from $(q_0,z_0)$ if there exists a run from $(q_0,z_0)$ to $(q_n,z_n)$.  
We write $\reach{(q_0,z_0)}{(q_n,z_n)}$ to denote the existence of such a run. 
Given a path $\pi$, we write $\RUN{(q_0,z_0)}{\pi}{(q_n,z_n)}$ if $\pi$ \emph{yields} a run from $(q_0,z_0)$ to $(q_n,z_n)$.

A path~$\pi$ has no hope to yield a run from $(q,z)$ if $z+\effect{\pi'}< 0$ for some prefix $\pi'$ of $\pi$.
We denote by $\drop{\pi}$ the maximum of $-\effect{\pi'}$ over all prefixes $\pi'$ of $\pi$, and call it the \emph{drop} of $\pi$. Intuitively, $\drop{\pi}$ is the smallest counter value $z\in \mathbb{N}$ such that $\pi$, when applied from $(q,z)$, remains nonnegative; note that hitting a guard is not a consideration here. 

We use the following standard operators: $\mathsf{Post}(c) \defeq \set{ c' \in \configs \mid c \xrightarrow{} c' }$ and $ \mathsf{Pre}(c) \defeq \set{ c' \in \configs \mid c'\xrightarrow{} c }$.
For $X \subseteq \configs$, we write $\Post X \defeq \bigcup_{c \in X} \Post c$ and $\Pre  X \defeq \bigcup_{c \in X} \Pre c$.
Also,
$ \posts^*(X)  \defeq \set{d \mid \exists c \in X \colon c \step{*} d}$ and $\pres^*(X) \defeq \set{c \mid \exists d \in X \colon c \step{*} d}$. 

For an OCA $\A = (Q, \transitions, \tau)$, we define the \emph{reverse of \A}\label{def:reverse} as $\rev{\A} \defeq (Q, \rev{\transitions}, \tau)$ where $(q,a,q')\in \rev{\transitions}$ if and only if $(q',-a,q)\in \transitions$.
Given configurations $c$ and $d$ and a path $\pi$ in $\A$, we have $\reach{c}{d}$ in $\A$ if and only if $\reach{d}{c}$ in $\rev{\A}$. 

\subparagraph*{The reachability problem.}
We consider the following decision problem.
\problemyy%
    {Reachability}%
    {An OCA $\VASS$ with equality and disequality tests, a valid initial configuration $\source$, and a valid target configuration $\target$.}%
    {Does $\reach{\source}{\target}$ hold?}
\smallskip
The model of OCA with disequality tests has been studied in~\cite{LechnerMOPW18} and~\cite{AlmagorCPS020}. 
The latter paper provides polynomial-time algorithms for the \emph{coverability problem}: ``given $\source$ and a state $q$, does there exist $z$ such that $\reach{\source}{(q,z)}$\,?'' and the related \emph{unboundedness problem}: ``is the set of configurations reachable from $\source$ infinite?''. The reachability problem, however, is \NP-hard even without tests, see \cref{fig:subset-sum} (Left).  

\subparagraph*{Equality tests.}
In the reachability problem in OCA with equality and disequality tests, the main technical challenge stems from disequality tests. Indeed, a state with an equality test only has one valid configuration hence need not be visited more than once.
\begin{mdframed}[style = attractattention] 
We now work with OCA with disequality tests only. We will discuss in \cref{subsec:equality-tests} how our techniques are affected by the addition of equality tests. 
\end{mdframed}

\begin{figure}
    \centering
    \input{figures/subset-sum}
    \quad
    \input{figures/chains}
    \caption{
	\newline
        \textbf{Left.} This OCA without tests is constructed from an instance of the subset sum problem $(a_1, \ldots, a_n, v)$; this is in fact how the reachability problem in OCA without tests is proved to be \compclass{NP}-hard in \protect\cite{HaaseKOW09}.
        Configuration $(t,0)$ can be reached from configuration $(s,0)$ whenever there exists a subset of $\set{a_1, \ldots, a_n}$ whose elements sum up to $v$.
        Note that all unlabelled transitions have update zero.
        The set of configurations reachable from $(s,0)$ can have size exponential in $n$, and its structure is unwieldy.
        \\
        \textbf{Right.} (For Section~\ref{sec:bg-oca}.) The named state $q$ belongs to $\posstates$ since there is a simple $q$-cycle with positive effect.
        There are six bounded chains of configurations at $q$.
        The disequality test $\neq 5$ bounds the counter values with residue $0$ modulo $5$, so $\set{(q, 0)}$ and $\set{(q, 5)}$ are bounded chains.
        The disequality test $\neq 30$ bounds the counter values with residue $3$ modulo $5$, so $\set{(q, 3), (q, 8), (q, 13), (q, 18), (q, 23),(q, 28)}$ is a bounded chain.
        The disequality test $\neq 15$ bounds the counter values with residue $2$ modulo $5$, so $\set{(q, 2), (q, 7), (q,12)}$ is a bounded chain.
    }
    \label{fig:subset-sum}
\end{figure}

%% file: figures/subset-sum.tex
\begin{tikzpicture}
	\node[circle, draw = black, line width = 0.4mm, minimum size = 4mm, inner sep = 0mm] (p1) at (0.4, 0) {\small$s$};
	\node[circle, draw = black, line width = 0.4mm, minimum size = 4mm, inner sep = 1mm] (p2) at (1.2, -0.5) {};
	\node[circle, draw = black, line width = 0.4mm, minimum size = 4mm, inner sep = 1mm] (p3) at (2, 0) {};
	\path[line width = 0.4mm, -{Stealth[width=1.8mm, length=2.2mm]}] (p1) edge (p2);
	\path[line width = 0.4mm, -{Stealth[width=1.8mm, length=2.2mm]}] (p2) edge (p3);
	\path[line width = 0.4mm, -{Stealth[width=1.8mm, length=2.2mm]}] (p1) edge[bend left = 10] node[above]{\small$+a_1$} (p3);

	\node[rotate = -22] (c1) at (3, -0.4) {\small$\cdots$};
	\path[line width = 0.4mm, -{Stealth[width=1.8mm, length=2.2mm]}] (p3) edge (c1);
	\node[rotate = 10] (c2) at (3, 0.3) {\small$\cdots$};
	\path[line width = 0.4mm, -{Stealth[width=1.8mm, length=2.2mm]}] (p3) edge (c2);

	\node[circle, draw = black, line width = 0.4mm, minimum size = 4mm, inner sep = 1mm] (q1) at (4.6, 0) {};
	\node[circle, draw = black, line width = 0.4mm, minimum size = 4mm, inner sep = 1mm] (q2) at (5.4, -0.5) {};
	\node[circle, draw = black, line width = 0.4mm, minimum size = 4mm, inner sep = 0mm] (q3) at (6.2, 0) {\small$r$};
	\path[line width = 0.4mm, -{Stealth[width=1.8mm, length=2.2mm]}] (q1) edge (q2);
	\path[line width = 0.4mm, -{Stealth[width=1.8mm, length=2.2mm]}] (q2) edge (q3);
	\path[line width = 0.4mm, -{Stealth[width=1.8mm, length=2.2mm]}] (q1) edge[bend left = 10] node[above]{\small$+a_n$} (q3);

	\node[rotate = 22] (d1) at (3.6, -0.4) {\small$\cdots$};
	\path[line width = 0.4mm, -{Stealth[width=1.8mm, length=2.2mm]}] (d1) edge (q1);
	\node[rotate = -10] (d2) at (3.6, 0.3) {\small$\cdots$};
	\path[line width = 0.4mm, -{Stealth[width=1.8mm, length=2.2mm]}] (d2) edge (q1);

	\node[circle, draw = black, line width = 0.4mm, minimum size = 4mm, inner sep = 0mm] (t) at (7.5, 0) {\small$t$};
	\path[line width = 0.4mm, -{Stealth[width=1.8mm, length=2.2mm]}] (q3) edge node[above]{\small$-v$}(t);
\end{tikzpicture}

%% file: figures/chains.tex
\begin{tikzpicture}[scale = 0.7]
	\node[circle, draw = black, line width = 0.4mm, minimum size = 6mm, inner sep = 1mm, label=right:{\small$\neq 5$}] (q) at (0, 0) {$q$};
	\node[circle, draw = black, line width = 0.4mm, minimum size = 6mm, inner sep = 1mm, label=right:{\small$\neq 30$}] (r) at (2, 1) {};
	\node[circle, draw = black, line width = 0.4mm, minimum size = 6mm, inner sep = 1mm, label=right:{\small$\neq 15$}] (s) at (2, -1) {};

	\path[line width = 0.4mm, -Stealth] (q) edge[bend left = 25] node[above left]{\small$+2$} (r);
	\path[line width = 0.4mm, -Stealth] (r) edge[bend left = 25] node[right]{\small$+1$} (s);
	\path[line width = 0.4mm, -Stealth] (s) edge[bend left = 25] node[below left]{\small$+2$} (q);

	\path[line width = 0.4mm, -Stealth] (-1, 0) edge (q);
\end{tikzpicture}

%% file: chapters/background.tex
In this section, we fix an OCA~$\VASS=(Q,\transitions,\guards)$ with disequality tests and two valid configurations~$\source$ and $\target$.

A configuration $c$ is \emph{bounded} when $\posts^*(c)$ is finite, and \emph{unbounded} otherwise. It is known, although far from trivial, that one can decide boundedness in polynomial time.

\begin{lemma}[see {\cite[Theorem 19]{AlmagorCPS020}}] \label{decide_boundedness}
    Given an OCA with disequality tests and a configuration~$c$, it is decidable in polynomial time whether $c$~is bounded or unbounded. 
\end{lemma}

A \emph{candidate run} is a run except that neither the nonnegativity condition nor the disequality tests are necessarily respected.
Formally, a candidate run is simply a sequence $(q_0,z_0), \ldots, (q_n,z_n)$ where all $(q_i, z_i) \in \states \times \ZZ$, $n \ge 0$, and such that there exist transitions $(q_{i-1}, a_i,q_{i})\in \transitions$ with $z_{i+1}=z_{i}+a_i$ for all $i\in \set{1, \ldots, n}$. 
We write $\candidate{(q_0,z_0)}{(q_n,z_n)}$.

An ingredient of the $\compclass{NP}$ upper bound for reachability in OCA \emph{without} disequality tests~\cite{HaaseKOW09} is establishing conditions under which a candidate run can be \emph{lifted} to a run. We adapt the argument to OCA with disequality tests.

\begin{restatable}{lemma}{sclifting} \label{unbounded_rev_unbounded}
    Let $\VASS$ be a strongly connected OCA with disequality tests.
    If $\source$ is unbounded in $\VASS$ and $\target$ is unbounded in $\rev{\VASS}$, then there is a run from $\source$ to $\target$ in $\VASS$ $(\source \xrightarrow{*} \target)$ if and only if there is a candidate run from $\source$ to $\target$ in $\VASS$ $(\candidate{\source}{\target})$.
\end{restatable}

The hypothesis that $\VASS$ is strongly connected is crucial. 
Indeed, if $\VASS$ is strongly connected and $\source = (s,v)$ is unbounded in $\VASS$, then there is a cycle of positive effect that, from $\source$, can be applied infinitely often to reach $(s,z)$ with $z$ arbitrarily large. 
If $\VASS$ is not strongly connected, it could be that the positive cycles that make $\source$ unbounded are in another SCC and that the set $\set{z \mid (s,v) \step{*} (s,z)}$ is finite. 
See~\cref{app:bg-oca} for a proof of~\cref{unbounded_rev_unbounded}.


Let $\posstates \subseteq Q$ be the set of states~$q\in Q$ such that there
exists a $q$-cycle~$\gamma$ with $\length{\gamma} \le |Q|$ and with $\effect{\gamma}>0$.
For each $q \in \posstates$, let $\poscycle q$ be such a $q$-cycle with minimal drop.
\textbf{We fix this choice for the remainder of the paper.}
We define
\begin{equation*} \label{def:confplus}
    \posconfigs \defeq \{(q,z) : q\in \posstates \text{ and } z\geq \drop{\poscycle q}\}.  
\end{equation*}

\begin{restatable}{lemma}{identify} \label{lem:define-canonical} \label{claim-membershipConfplus}
    There is a polynomial-time algorithm to identify $\posstates$ and to choose cycles $\poscycle q$ for all $q \in \posstates$. 
    Moreover, membership in $\posconfigs$ can be decided in polynomial time.  
\end{restatable}

\begin{remark}
    Our choice of $\posstates$ differs slightly from the definition found in~\cite{AlmagorCPS020}: we use short cycles ($\length{\gamma} \leq |Q|$) rather than simple cycles.
    For simple cycles, the ability to compute, in polynomial time, the minimal drop of a positive-effect simple $q$-cycle (for each $q \in Q$) is not justified in~\cite{AlmagorCPS020}.
    In fact, in~\cref{app:finding-simple-cycles-is-hard}, we prove that deciding, for a given OCA without tests $\Aa$ and a given state $q$, whether there exists a positive-effect simple $q$-cycle in $\Aa$ is an \NP-complete problem.
    However, all constructions and arguments of~\cite{AlmagorCPS020} appear to be insensitive to the replacement of `simple cycles' by `short cycles'.
    As a result, we can still use polynomial-time algorithms for coverability and for unboundedness in OCA with disequality tests (1-VASS with disequality tests).
\end{remark}

The set of all $(q,z) \in \posconfigs$
can be partitioned into \emph{$q$-chains}.
For each $q \in \posstates$, let $\posconfigs(q) =  (\{q\} \times \NN) \cap \posconfigs$.
A $q$-chain $\chain$ is a maximal non-empty subset $\achain \sset \posconfigs(q)$ such that, for every two distinct $\aconfig, \aconfig' \in \achain$, either $\aconfig$ is reachable from $\aconfig'$ by iterating $\poscycle{q}$, or vice versa.  
In other words, $\achain$ is a non-empty minimal subset of $\posconfigs(q)$ (with respect to set inclusion) such that, for all $\aconfig \in \achain$ and all $\aconfig' \in \configs$, if $\aconfig \step{\poscycle{q}} \aconfig'$ or $\aconfig' \step{\poscycle{q}} \aconfig$ then $\aconfig' \in \achain$.

A  $q$-chain is \emph{bounded} if it is a finite set, otherwise it is \emph{unbounded}. 
Note that configurations in unbounded chains are all themselves unbounded, but configurations in bounded chains need not be bounded (they may be unbounded). 
Because the number of disequality guards that a cycle $\gamma_q$ may encounter is small, so is the total number of bounded chains.

\begin{restatable}[see~{\cite[Remark 6]{AlmagorCPS020}}]{lemma}{chaincount} \label{rem-number-bounded-chains}
    There are at most $2\size{Q}^2$ bounded chains.
\end{restatable}

Given a chain $C$, the counter values~$z$ of every $(q,z) \in C$ have the same remainder modulo~$\effect{\poscycle q}$.
Henceforth, a bounded $q$-chain can be described as 
$[\ell,u] \, \cap \, (r+ \effect{\poscycle q} \cdot \mathbb{N})$
where $[\ell,u]$ is an interval of nonnegative integers and $r+ \effect{\poscycle q} \cdot \mathbb{N}$ is
 an arithmetic progression with initial term~$r$ and difference~$\effect{\poscycle q}$. 
Since the OCA~$\VASS$ is encoded in binary, the values of $l$, $u$, $r$, and $|\poscycle{q}|$ may be exponential in the size of $\VASS$.
See~\cref{fig:subset-sum} (Left) for an example.

%% file: chapters/pessimistic.tex
In this section, we exhibit a family run of runs, namely pessimistic runs, that are guaranteed to admit an \NP certificate. 
This will already enable us to prove, in Section~\ref{sec:strongly-connected-OCA}, that the reachability problem is in \pitwop in the special case where the OCA is strongly connected. 

Let $\VASS$ be an OCA with disequality tests.
We call a run of~$\VASS$ \emph{pessimistic} if none of its configurations are in $\posconfigs$, except possibly the first one. 
Of course, some pessimistic runs may be exponentially long relative to the size of $\VASS$; however, we provide a way to handle them.
For $S \subseteq \posconfigs$, we write $\poststarpessimistic{S}$ for the set of configurations reachable from $S$ using only pessimistic runs. 
In particular, $S \subseteq \poststarpessimistic{S}$.

Consider the following decision problem:
\problemyy%
{Pessimistic Reachability}%
{An OCA $\VASS$ with disequality tests, and two configurations $\source$ and $\target$.}
{Is there a pessimistic run from $\source$ to $\target$ in $\VASS$?}
\smallskip

Pessimistic runs turn out to be very handy, not least because we can adapt an existing `flow' technique~\cite{HaaseKOW09} to decide pessimistic reachability.

\begin{lemma}
	\label{certificate_pessimistic_runs_cc}
	The pessimistic reachability problem is in \textup{\NP}.
\end{lemma}

In a nutshell, the idea~\cite{HaaseKOW09} is to guess how many times the run traverses each transition.
The guessed numbers are subject to polynomial-time checkable balance and connectivity conditions, akin to, e.g.,~\cite{SeidlSMH04}.
However, we cannot check whether the (possibly very long) run constructed from the flow violates disequality constraints,
so the technique cannot be applied directly.

Our solution (detailed in \cref{app:pessimistic-np}) uses the pessimism of the run.
Let $x \neq g$ be a guard on state~$q$.
We split the run in two:
in the first part, all visits to~$q$ are above~$g$; then the run \emph{jumps} the guard so that, in the second part, all visits are below~$g$.
This way,
with at most $|Q|$ splits,
we can reduce the problem to the case in which the run does not jump \emph{any} disequality guard (always staying above or below each of them).

%% file: chapters/strongly-connected.tex
In this section, for pedagogical purposes, we study the particular case where the OCA is strongly connected. The case with multiple SCCs presented in \cref{sec:several-SCCs} is more technical but relies on the same key idea. 

\begin{theorem}
	\label{thm:strong}
	The reachability problem for strongly connected OCA with disequality tests
	belongs to the complexity class \pitwop.
\end{theorem}

We sketch the proof of Theorem~\ref{thm:strong} below.
Throughout the section, we fix a strongly connected OCA with disequality tests $\VASS$ and two configurations $\source$ and $\target$, and we are interested in whether $\source \step{*} \target$.

\subsection{Ruling out the Unbounded Case}
\label{subsec:unbounded-sc}
By~\Cref{decide_boundedness}, given an instance of reachability, we can check in polynomial time whether $\source$ is unbounded in $\VASS$ and $\target$ is unbounded in $\rev{\VASS}$.
If both are true, then, by \Cref{unbounded_rev_unbounded}, it suffices to determine whether there exists a candidate run from $\source$ to $\target$.
The existence of a candidate run can be decided in \NP (e.g., using integer linear programming, see~\cite{BT76}). 
This case will not affect our complexity result because $\NP \sset \pitwop$.
Thus, without loss of generality, we may assume that $\source$ is bounded in $\VASS$ or $\target$ is bounded in $\rev{\VASS}$. Moreover,
if $\target$ is bounded in $\rev{\VASS}$, we symmetrically work with $\rev{\VASS}$ instead of $\VASS$.

\begin{mdframed}[style = attractattention]
In the remainder of this section, we assume that $\source$ is bounded in $\VASS$.
\end{mdframed}

\subsection{Inductive Invariants in the Bounded Case}
\label{subsec:strongly-connected-invariants}
We will show that
$\notreach{\source}{\target}$ if and only if there exists a certificate of a particular shape witnessing this non-reachability.
This certificate takes the form of an inductive invariant separating $\source$ and $\target$. 
The exact set of configurations comprising this inductive invariant is unwieldy, so we concentrate on its \emph{core} instead. 
This set of core configurations admits a short representation, as follows.

We call an \emph{arithmetic progression} on state $q \in \states$ a set of configurations $\set{(q,v) \mid \ell \leq v \leq L \land \exists k \in \nats, v= k p + s}$ with $p,s,\ell,L \in \nats$.
An arithmetic progression can be specified by writing $q$ and the numbers $p, s, \ell, L$.
A set of configurations \emph{has a concise description} if it is a union of at most $2 \size{Q}^2 + 1$ arithmetic progressions whose configurations have counter value bounded by $2 \size{Q} \cdot \norm{\transitions} \cdot \norm{\diseqtests}$.
Such a set can be described in polynomial space. 

The set of all configurations in bounded chains has a concise description. This also holds for the set $R$ of all \emph{reachable} configurations in bounded chains: indeed, if a configuration of a chain can be reached, the same is true for all configurations above in the same chain. Because $\source$ is bounded, unbounded chains cannot be reached, and this $R$ is in fact the set of reachable configurations in all chains. 
(Observe that runs that reach configurations from $R$ may well visit the complement of $\posconfigs$.)

\begin{lemma}
	\label{reachable_in_chain_reasonable}
	The set $R$ of reachable configurations in $\posconfigs$ has a concise description.
\end{lemma}

Intuitively, $R$ is our desired `core invariant', and the desired invariant is the set $\poststarpessimistic{R}$. 
However, when given a set $I$, it is not easy to check whether $I$ is actually equal to $R$. Instead,
the following theorem defines possible invariant cores by $3$~conditions.

Conditions involving $\source$ and $\target$ are self-explanatory.
Set inclusion
$\Post{\poststarpessimistic{I}} \subseteq \poststarpessimistic{I}$ would express inductiveness
(closure of the set under $\Post{\cdot}$).
However, verifying this condition is computationally expensive,
and we replace it with a version that `focuses' on the core only, and thus has $I$ rather than $\poststarpessimistic{I}$
on the right-hand side.

\begin{theorem}\label{unreachability_invariant_better}
	Suppose $\source$ is bounded in $\VASS$. 
       Then $\notreach{\source}{\target}$ if and only if there exists a set $I \subseteq \posconfigs \cup \set{\source}$ with concise description such that:
		\begin{enumerate}[label=\textup{\textsf{(Cond}\textsf{\arabic*}\textsf{)}},ref=\textsf{(Cond}\textsf{\arabic*}\textsf{)}, leftmargin=50pt]
		\item \label{item:inv-sc-source} $\source \in I$,
		\item \label{item:inv-sc-target} $\target \notin \poststarpessimistic{I}$, and
		\item \label{item:inv-sc-invariant} $\Post{\poststarpessimistic{I}} \cap \posconfigs \subseteq I$.
	\end{enumerate}
\end{theorem}
\begin{proof}
	First, assume that there is such a set $I$.
	Because $\target \notin \poststarpessimistic{I}$ by \ref{item:inv-sc-target}, it suffices to prove that $\posts^*(\source) \subseteq \poststarpessimistic{I}$.
	We proceed by induction on the length of the run from $\source$ to $c \in \posts^*(\source)$. The base of induction is \ref{item:inv-sc-source}. Assume that we have $d \in \poststarpessimistic{I}$ and $d \step{} c$. If $c \notin \posconfigs$ then we have a pessimistic run from $I$ to $c$, so $c \in \poststarpessimistic{I}$. If $c \in \posconfigs$ then $c \in \Post{\poststarpessimistic{I}} \cap \posconfigs$, hence $c \in I$ by \ref{item:inv-sc-invariant}. 
    For the other direction,
	assume that $\target$ is not reachable from $\source$. Let $I \defeq R \cup \{\source\}$; by \cref{reachable_in_chain_reasonable}, $I$ has a concise description. \ref{item:inv-sc-source} and \ref{item:inv-sc-target} are trivially satisfied. 
	Moreover, $\Post{\poststarpessimistic{I}} \cap \posconfigs \subseteq \posts^*(\source) \cap \posconfigs \subseteq I$, hence \ref{item:inv-sc-invariant} is satisfied.
\end{proof}

\subsection{The Complexity of Reachability in Strongly Connected OCA}
\label{subsec:strongly-connected-complexity}
We now prove that reachability is in \pitwop by, equivalently, proving that \emph{non}-reachability is in \sigmatwop.
We refer the reader to Appendix~\ref{app:def-ph} for definitions of these complexity classes.
For our purpose, the following informal description suffices. A problem is in \sigmatwop whenever this problem is solvable in non-deterministic polynomial time by a Turing machine which has access to an oracle for some \NP-complete problem. The oracle is a black box that may provide the answer to any problem in \NP (and therefore to any problem in \compclass{coNP}).  

As argued in \cref{subsec:unbounded-sc}, we assume with no loss of generality that $\source$ is bounded.
By \cref{unreachability_invariant_better}, we have 
$\notreach{\source}{\target}$ if and only if there exists $I$ satisfying the three conditions \ref{item:inv-sc-source}, \ref{item:inv-sc-target}, and \ref{item:inv-sc-invariant}.
Moreover, by the same theorem, $I$ can be assumed to have a concise description.
Thus, we can guess such a set $I$ in non-deterministic polynomial time. It remains to prove that the verification that a set $I$ satisfies the three conditions can be performed using an \NP oracle. To this end, we prove that this verification is a \coNP problem. 
Indeed, $I$ does \emph{not} satisfy the three conditions when: 
\begin{itemize}
	\item either $\source \not\in I$ (which can be checked efficiently),
	\item or $\target \in \poststarpessimistic{I}$ (this is when there is a \emph{small} configuration $c$ such that $c \in I$ and $\target \in \pposts{c}$),
	\item or there are some \emph{small} configurations $c$ and $d$ such that $c \in I$, $d \in \poststarpessimistic{c}$, and some successor of $d$ belongs	to $\posconfigs$ but not to $I$.
\end{itemize} %
The adjective \emph{small} should here be understood as ``bounded by an exponential in the size of $\Aa$, $\source$, and $\target$''.
In fact, it is fairly easy to obtain an exponential bound on configurations to consider, see \cref{app:exp_bounds}.
Thanks to~\cref{certificate_pessimistic_runs_cc}, verification of both whether there is a $c \in I$, and $e \in \Post{d}$ such that $\target \in \pposts{c}$ and whether there exist $c \in I$, $d \in \poststarpessimistic{c}$ such that $e \notin I$ are in \NP.
Since membership in $\posconfigs$ can be checked in polynomial time by~\cref{claim-membershipConfplus}, 
the entire third condition is also an \NP condition.
This completes the proof of~\cref{thm:strong}.

%% file: chapters/leaky.tex
In this section, we extend the techniques from \cref{sec:strongly-connected-OCA} to the general case in which the OCA is not assumed to be strongly connected. 
We fix an OCA \A with disequality tests and two configurations $\source$ and $\target$. 

We first highlight why the techniques developed above do not apply to this general case.
In \cref{sec:strongly-connected-OCA}, the hypothesis that \A is strongly connected was necessary for the application of \cref{unbounded_rev_unbounded}. When \A is not strongly connected, knowing that $\source$ is unbounded is no longer satisfactory. Indeed, it no longer implies the existence of a positive cycle involving its state, as the positive cycle allowing us to pump up could be in another SCC. We need to be able to specify whether a configuration is unbounded \emph{within its own SCC} or not.

\subsection{Locally Bounded Configurations and Runs}
\label{sec:locally-bounded}

\subparagraph*{Locally bounded configurations.} 
Given a SCC $S$ of \A, we denote by $\A_S$ the automaton obtained when restricting \A to states and transitions within $S$. A configuration $c$ is \emph{locally bounded} if
$c$ is bounded in $\A_S$ where $S$ is the SCC of $c$. 
We denote by $L$ the set of all locally bounded configurations (and by $\rev{L}$ in $\rev{\A}$).
Configurations that are not locally bounded are referred to as
\emph{locally unbounded}.
We generalise the lifting technique from~\cref{unbounded_rev_unbounded}.
The proof of the following lemma can be found in~\cref{app:locally-bounded-proofs}.

\begin{restatable}[Lifting]{lemma}{lifting} \label{unbounded_rev_unbounded_global}
	For all $c \notin L$  and $d \notin \rev{L}$, we have $c \step{*} d$ if and only if $\candidate{c}{d}$.
\end{restatable}

\subparagraph*{Locally bounded runs.}
A run $c\xrightarrow{\pi}d$ is said to be \emph{locally bounded} if all configurations visited by the run are locally bounded. We denote such a run by $c\xrightarrow[L]{\pi}d$, and denote its existence by $c\xrightarrow[L]{*}d$. Notice that a locally bounded run may go through several SCCs. Moreover, a run starting from a locally bounded configuration is not always locally bounded: once it goes to a new SCC, it may visit configurations that are not locally bounded. 
We define the locally bounded counterpart $\bppostsymb$
of the pessimistic post-star operator: $d \in \bpposts{c}$ if there is a pessimistic and locally bounded run from $c$ to $d$.
We extend this definition to sets of configurations $X$ in the usual way.
Dually, we also define, for every set of configurations $X$, the set $\boprestar{X}$ as the same notion but in the reverse OCA $\rev{\A}$. 
We extend~\cref{certificate_pessimistic_runs_cc} to these new operators.

\begin{restatable}{lemma}{lpostnp} \label{lem:lpost-np}
	Given $c, d \in \configs$, deciding whether $d \in \bpposts c$ is in \NP.
\end{restatable}
\begin{proof}[Proof sketch]
	We split the run on its transitions between SCCs. 
	We apply~\cref{certificate_pessimistic_runs_cc} on the portions remaining in one SCC.
	Since the run is pessimistic, we can bound all the counter values in it.
	The run is locally bounded when the first configuration visited in each SCC is locally bounded, which is checked using~\cref{decide_boundedness}. 
	The full proof can be found in~\cref{app:locally-bounded-proofs}.
\end{proof}

\subsection{Leaky Invariants}
Unlike in the strongly-connected case, a single invariant construction is not sufficient for our needs. Indeed, if $\source$ is locally bounded but unbounded, then one could imagine the invariant technique from \cref{unreachability_invariant_better} applied to the SCC $S$ of $\source$, but then this invariant would not apply to other SCCs. 
For example, there could be runs that are locally bounded in the SCC $S_{\source}$ of $\source$ but not in the SCC $S_{\target}$ of $\target$, making the invariant inapplicable. 
Instead, assuming that $\target$ is locally bounded in $\rev{\A}$, one may consider in the SCC of $\target$ an invariant constructed in the reverse automaton $\rev{\A}$. 
We therefore employ a pair of invariants, one for $\Aa$ (the \emph{forward invariant}) and another one for its reverse~$\rev{\Aa}$ (the \emph{backward invariant}).
The two invariants will induce two sets of configurations that, in a negative instance of the reachability problem, separate the source and target.

The following lemma will allow us to avoid treating $\source$ and $\target$ separately.
The set $\rev{\posconfigs}$ is defined as the counterpart of $\posconfigs$ in $\rev{\A}$.
\begin{lemma}
\label{source_target_simplification}
We may assume that $\source\in \posconfigs \cap L$ and $\target \in \rev{\posconfigs} \cap \rev{L}$.
\end{lemma}  

We now define our notion of a leaky invariant. 
As in \Cref{sec:strongly-connected-OCA}, we represent the invariants using \emph{core} sets of configurations that can be succinctly described, denoted by $I$ and $J$.
Our invariants must be inductive in the following weak sense:

\begin{condition}\label{con:inductive}
	Let $I \subseteq \posconfigs \cap L$ and $J \subseteq \rev{\posconfigs}\cap \rev{L}$ be sets of configurations. \\
	The pair $(I, J)$ is \textbf{inductive} if

	\vspace{0.1in}\noindent
	\begin{tabular}{l l}
	\quad \multirow{2}{*}{\ind} \quad\quad 
	& $\Post{\bpposts{I}} \cap \posconfigs \cap L \sset I$ \quad and \\[0.05in]
	& $\Pre{\bopres{J}} \cap \rev{\posconfigs} \cap \rev{L} \sset J $. 
	\end{tabular}
\end{condition}

Notice that $I$ and $J$ play symmetric roles in $\A$ and $\rev{\A}$.
We now provide some intuition for the (forward) inductive condition for $I$.
The set $I$ only contains configurations from $\posconfigs \cap L$, because the set $\posconfigs \cap L$ has a regular structure thanks to bounded chains. The set $I$ is, again, only the \emph{core} of the invariant. The full invariant\footnote{Our invariant is $\pposts{I} \cup \Post{\pposts{I}}$ but the operator that appears in \cref{con:inductive} is $\bpposts{\cdot}$. In \cref{app:discussion-operators}, we discuss the issues encountered if $\pposts{\cdot} \cup \Post{\pposts{\cdot}}$ was used in \cref{con:inductive}.} is $\pposts{I} \cup \Post{\pposts{I}}$, but this set is not easily described  (see \cref{tractable}). This explains why we use the composition $\Post{\bpposts{\cdot}}$ instead of the single-step $\Post{\cdot}$ operator traditionally used to define inductiveness. 

\begin{remark} \label{rem:leaky}
We refer to our invariants as \emph{leaky}, because they are not inductive in the traditional sense. Indeed, our invariants are `focused' on locally bounded configurations, and can be escaped by transitions to locally unbounded configurations. This leak may, however, only happen with transitions going from one SCC to another. 
\end{remark}

\begin{condition}\label{con:separator}
	Let $\finv, \binv \subseteq \configs$ be sets of configurations. \\
	The pair $(\finv, \binv)$ is a \textbf{separator} if, for all $c \in \finv$ and $d \in \binv$,
	\begin{enumerate}[label=\textup{\textsf{(Sep}\textsf{\arabic*}\textsf{)}},ref=\textsf{(Sep}\textsf{\arabic*}\textsf{)}, leftmargin=50pt]
	\item \label{sep1} $c \unstep d$; and 
	\item \label{sep2} if $c \notin L$ and $d \notin \rev{L}$, then $\notcandidate{c}{d}$. 
	\end{enumerate}
\end{condition}

Firstly, \ref{sep1} will forbid $\finv$ and $\binv$ from being connected by a single step.
Secondly, \ref{sep2} will forbid connection between $\finv$ and $\binv$ using the lifting technique of \Cref{unbounded_rev_unbounded_global}. This does not, in general, prevent the existence of runs from $\finv$ to $\binv$; it will do so, however, for our leaky invariants that combine Conditions~\ref{con:inductive} and~\ref{con:separator}. 

\begin{mynewdefinition} \label{def:leaky-invariant}
	Let $I \subseteq \posconfigs \cap L$ and $J \subseteq \rev{\posconfigs} \cap \rev{L}$. 
    Consider the sets 
    \begin{align*}
    	\finv & \coloneqq \pposts{I} \cup \Post{\pposts{I}} \text{ and }\\
    	\binv & \coloneqq \opres{J} \cup \Pre{\opres{J}}.
    \end{align*}
    We call the pair $(I,J)$ a \emph{non-reachability witness} for $\source$ and $\target$
    if $(I, J)$ is inductive, $(\finv, \binv)$ is a separator, $\source \in I$, and $\target \in J$. 
\end{mynewdefinition}

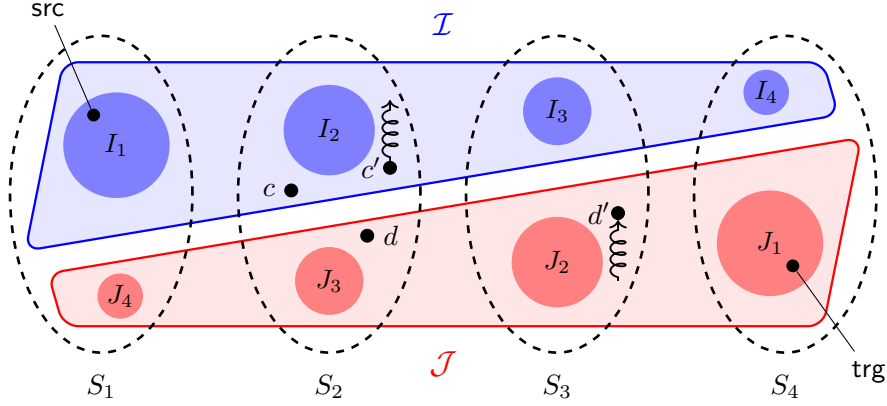
\begin{figure}
	\centering
	\input{figures/abstract-sets}
	\caption{Core inductive sets and the separator they induce.
	The core of the forward leaky invariant is $I = I_1 \cup I_2 \cup I_3 \cup I_4$ (the blue circular sets) and the core of the backward leaky invariant is $J = J_1 \cup J_2 \cup J_3 \cup J_4$ (the red circular sets). 
	The induced separator $(\finv, \binv)$ is shown as blue and red rounded quadrilaterals containing the core sets.
	Notice that $\source \in I$ and $\target \in J$.
	The upwards coiled arrow from $c'$ represents that $c'$ is locally unbounded in the SCC $S_2$ and the downwards coiled arrow from $d'$ represents that $d'$ is locally unbounded in the SCC $\rev{S_3}$.
	Note also the separator conditions:
	Condition~\ref{con:separator}~\ref{sep1} means that configurations $c \in \finv$ and $d \in \binv$ cannot reach one another by one transition, so $c \unstep{} d$;
	Condition~\ref{con:separator}~\ref{sep2} means that unbounded configurations $c' \in \finv$ and $d' \in \binv$ cannot reach one another via a candidate run, so $\notcandidate{c}{d}$.
	}
	\label{fig:abstract-sets}
\end{figure}

The pair of sets $(I, J)$ forms the \emph{core} of the invariant, namely $I$ represents the \emph{forward leaky invariant} and $J$ represents the \emph{backward leaky invariant}. 
In this case we also say that $(I, J)$ \emph{induces} the separator $(\finv, \binv)$.
A visualisation of a pair $(I,J)$ and its induced separator $(\finv, \binv)$ can be seen in~\cref{fig:abstract-sets}.
A helpful intuition is that $\finv$ is approximately $\pposts{I}$ (and similarly $\binv$ is approximately
$\opres{J}$). 
One additional step of $\Post{\cdot}$ (and $\Pre{\cdot}$, respectively) ensures that the ``outer boundary'' of this closure should also be included in the set.

\begin{remark}
\label{tractable}
	As in~\cref{sec:strongly-connected-OCA},
	our representation of invariants refers to their core only,
	i.e., the pair $(I, J)$.
	The example in Fig.~\ref{fig:subset-sum} (left) demonstrates
	that the set $\pposts{I} \cup \Post{\pposts{I}}$ does not always have a tractable description.
	The set of all possible sums of subsets has no convenient description, therefore we want it to be captured by $\pposts{I} \cup \Post{\pposts{I}}$ only and not by $I$ itself.
\end{remark}

\begin{restatable}{theorem}{leakyinvariant}
\label{lem:leaky-invariant}
	In an OCA $\VASS$ with disequality tests, $\target$ is \emph{not} reachable from $\source$ if and only if there exists a non-reachability witness. Moreover, in this case, there is always a non-reachability witness with a concise description. 
\end{restatable} 

In Section~\ref{sec:leaky:sound-and-complete}
we define `perfect cores', which we use in~\cref{sec:wrapping-up} to sketch a proof of~\cref{lem:leaky-invariant},
details being relegated to~\cref{app:wrapping-up-proofs}. 

\subsection{Perfect Cores}
\label{sec:leaky:sound-and-complete}

Condition \ind on the core of leaky invariants captures a weak inductiveness property, which is central to our approach.
We will now discuss two features of this condition that are used in the proof of \Cref{lem:leaky-invariant}.

Our conditions capture a specific invariant,
which we now define.
Consider the set
\begin{equation*}
	B \eqdef \set{ c \in \configs : \source \xrightarrow[L]{*} c }
\end{equation*}
In words, $B$ contains all configurations reachable from $\source$ using locally bounded runs. In line with ideas from Section~\ref{sec:strongly-connected-OCA}, we do not want to store $B$ entirely, so we will %
restrict the core to configurations in bounded chains.
We call \emph{perfect core} the set $B \cap \posconfigs$; the term \emph{perfect} is motivated by the fact that Definition~\ref{def:leaky-invariant} \emph{aims} to capture this set exactly.
Similarly, let $\rev{B} \coloneqq \set{ c \in \configs : c \xrightarrow[\rev{L}]{*} \target}$. The \emph{perfect core} in the reverse automaton is $\rev{B} \cap \rev{\posconfigs}$.

The two features of \ind are summarised in the following two lemmas.
We use the words ``sound'' and ``complete'' to characterise the relationship between Condition~\ref{con:inductive}  (as part of Definition~\ref{def:leaky-invariant}) and the perfect cores defined above.
Completeness expresses that in \emph{every instance} of non-reachability, the perfect cores defined above induce a non-reachability invariant.
Conversely, soundness states that \emph{every invariant} must contain all configurations from the perfect cores.
(Thus, the perfect core are the smallest possible invariants.)
The full proofs of soundness and completeness can be found in \cref{app:sound-complete}.

\begin{restatable}[Soundness]{lemma}{soundness} \label{lem:soundness}
	For all $I \subseteq \posconfigs \cap L$ and $J \subseteq \rev{\posconfigs} \cap \rev{L}$ such that $\source \in I$ and $\target \in J$, if $(I, J)$ is inductive (Condition~\ref{con:inductive}), then $B \cap \posconfigs \sset I \text{ and } \rev{B}\cap\rev{\posconfigs} \sset J$.
\end{restatable} 
\begin{proof}[Proof sketch]
\cref{con:inductive} for $I$ gives $\Post{\bpposts{I}} \cap \posconfigs \cap L \sset I$. Let $c \in B \cap \posconfigs$ be a configuration of the perfect core. Thus, $\source$ reaches $c$ by a locally bounded run. It is not always true that $c \in \Post{\bpposts{\source}}$ because this run does not have to be pessimistic: it may observe configurations in $\posconfigs$. We prove by induction that all configurations in $\posconfigs$ along the run are in $I$, using the property that $\Post{\bpposts{I}} \cap \posconfigs \cap L \sset I$ once for each such configuration; this eventually proves that $c \in I$. The proof is analogous for $J$. 
\end{proof}

\begin{restatable}[Completeness]{lemma}{completeness} \label{lem:completeness}
	If $I = B \cap \posconfigs$ and $J = \rev{B} \cap \rev{\posconfigs}$, then $(I, J)$ is inductive (Condition~\ref{con:inductive}).
\end{restatable}
\begin{proof}[Proof sketch]
We prove that $\Post{\bpposts{ B \cap \posconfigs }} \cap \posconfigs \cap L \sset B \cap \posconfigs$.
Let $c \in \posconfigs$ be locally bounded and belong to $\Post{\bpposts{ B \cap \posconfigs}}$. All configurations in $\bpposts{B \cap \posconfigs}$ are in $B$ by the definition of $B$, so $c$ can be reached in one step from a configuration $d \in B$. By definition, $d$ is reachable from $\source$ with a locally bounded run; since $c$ is itself locally bounded, this is also true for $c$, and so $c \in B$. The case of $J$ is similar.
\end{proof}

\subsection{Non-reachability Witnesses and Their Complexity}
\label{sec:wrapping-up}
 
We are ready to prove our main results.
Here, we will provide proof sketches, the full proofs can be found in~\cref{app:wrapping-up-proofs}.

\leakyinvariant*
\begin{proof}[Proof sketch]
	First, if $\notreach{\source}{\target}$ then the perfect cores $I = B \cap \posconfigs$ and $J = \rev{B} \cap \rev{\posconfigs}$ form a non-reachability witness.
	Indeed, by~\cref{lem:completeness}, $(I, J)$ is inductive. 
	Moreover, the induced $\finv = \pposts{I} \cup \Post{\pposts{I}}$ and $\binv = \opres{J} \cup \Pre{\opres{J}}$ form a separator. 
	We have $\finv \subseteq \posts^*(\source)$ and $J \subseteq \pres^*(\target)$, proving Condition~\ref{con:separator} \ref{sep1}. If Condition~\ref{con:separator} \ref{sep2} fails, \Cref{unbounded_rev_unbounded_global} yields a contradiction.
	Moreover, $I$ and $J$ have a concise description thanks to bounded chains.

	Conversely, suppose there is a non-reachability witness $(I, J)$.
	Assume for the sake of contradiction that $\source \step{*} \target$. 
	By~\Cref{lem:soundness}, since $(I, J)$ is inductive, $B \cap \posconfigs \sset I$ and $\rev{B} \cap \rev{\posconfigs} \sset J$.
	Consider a run from $\source$ to $\target$. 
	It must leave $\finv = \pposts{I} \cup \Post{\pposts{I}}$ therefore it visits locally unbounded configurations. Let $c$ be the first such configuration visited.
	Similarly, let $d$ be the last visited configuration that is locally unbounded in $\rev{\A}$.
	First, if $c$ occurs before $d$, then Condition~\ref{con:separator} \ref{sep2} is violated. 
	Second, if $c$ occurs after $d$, then there is an overlap in the runs from $\source$ to $c$ and from $d$ to $\target$.
	The overlap must be in $\finv \cap \binv$, leading to a violation of Condition~\ref{con:separator} \ref{sep1}. 
\end{proof}

\begin{restatable}{theorem}{complexity}
	\label{thm:complexity}
	The reachability problem for OCA with disequality tests belongs to the complexity class \pitwop.
\end{restatable} 
\begin{proof}[Proof sketch.]
	We use \cref{lem:leaky-invariant}, deciding the existence of a non-reachability witness in \sigmatwop. Recall that \sigmatwop is introduced in \cref{subsec:strongly-connected-complexity}.
	Let a pair $(I, J)$ be given; we show that a violation of the conditions for being a non-reachability witness can be checked in \compclass{NP}.
	\begin{itemize}
		\item
			One can check in polynomial time whether $\source \in I$ and $\target \in J$.
		\item
			Violation of Condition~\ref{con:inductive}~\ind is an \NP property.
			Indeed, this follows because membership in $\bpposts{\cdot}$ is in \NP (by \Cref{lem:lpost-np}) and membership in $\posconfigs$ and $L$ is polynomial-time checkable (by \Cref{claim-membershipConfplus} and \Cref{decide_boundedness}, respectively).
		\item
			Violation of Condition~\ref{con:separator}~\ref{sep1} is in \NP, because $\finv \coloneqq \pposts{I} \cup \Post{\pposts{I}}$, $\binv \coloneqq \opres{J} \cup \Pre{\opres{J}}$, and membership of given configurations in the pessimistic post-star (optimistic pre-star, respectively)	is in \NP by~\cref{certificate_pessimistic_runs_cc}.
			This assumes that we have an exponential bound on relevant configurations, which is guaranteed by~\cref{pessimistic_run_bounds} in the appendix.
		\item
			To check violation of Condition~\ref{con:separator}~\ref{sep2} in \NP, we again use~\cref{decide_boundedness} for $L$, as well as the fact that	the existence of a candidate run is in \NP (by integer programming).
	\qedhere
	\end{itemize}
\end{proof}

\subsection{Adding Equality Tests}
\label{subsec:equality-tests}

The previous techniques have been developed for OCA with disequality tests only. 
In particular, the lifting argument of~\cref{unbounded_rev_unbounded_global} does not hold in the presence of equality tests: candidate runs that visit a state with an equality test cannot be lifted to greater counter values. 
However, at the cost of increasing the complexity, one can handle equality tests. 

Complexity class \deltathreep consists of decision problems solvable in polynomial time with access to an $\NP^{\NP}$ oracle
(which can solve \sigmatwop problems in one step).

\begin{corollary}
	\label{equality-delta3}
	The reachability problem for OCA with equality and disequality tests
	belongs to the complexity class \deltathreep.
\end{corollary}
\begin{proof}
Let $\VASS$ be such an OCA with tests, and $\source$ and $\target$ two configurations. 
Denote by $\configs_=$ the set of valid configurations at states with equality tests; $\size{\configs_=}$ does not exceed the number of states in $\VASS$.
Consider the OCA with disequality tests $\VASS'$ that is obtained by deleting all states with equality tests (and incident transitions) from $\VASS$. 
By \cref{thm:complexity}, with an $\sigmatwop$ oracle we can build a graph with vertex set $\configs_= \cup \set{\source, \target}$ and edge set $\{(c,d) \mid \text{$\reach{c}{d}$ in $\VASS'$}\}$.
Depth-first search in this graph for a path from $\source$ to $\target$ takes polynomial time.
\end{proof}

%% file: figures/abstract-sets.tex
\begin{tikzpicture}
	\draw[rounded corners=2mm, draw = blue, line width = 0.3mm, fill=blue!10] (0,0)--(-0.5,-2.5)--(10.2,-0.7)--(10,0)--cycle;
	\node at (5, 0.5) {\Large\textcolor{blue}{$\finv$}};
	\draw[rounded corners=2mm, draw = red, line width = 0.3mm, fill=red!10] (-0.2,-2.8)--(0,-3.5)--(10,-3.5)--(10.5,-1)--cycle;
	\node at (5, -4) {\Large\textcolor{red}{$\binv$}};

	\draw[draw = black, line width = 0.35mm, dashed] (0.5,-1.75) ellipse (1.2cm and 2.1cm);
	\node at (0.5, -4.3) {\large$S_1$};
	\draw[draw = black, line width = 0.35mm, dashed] (3.5,-1.75) ellipse (1.2cm and 2.1cm);
	\node at (3.5, -4.3) {\large$S_2$};
	\draw[draw = black, line width = 0.35mm, dashed] (6.5,-1.75) ellipse (1.2cm and 2.1cm);
	\node at (6.5, -4.3) {\large$S_3$};
	\draw[draw = black, line width = 0.35mm, dashed] (9.5,-1.75) ellipse (1.2cm and 2.1cm);
	\node at (9.5, -4.3) {\large$S_4$};

	\node[circle, fill = red!50, minimum size = 6mm] at (0.75, -3.1) {};
	\node at (0.75, -3.1) {$J_4$};
	\node[circle, fill = red!50, minimum size = 9mm] at (3.5, -2.9) {$J_3$};
	\node[circle, fill = red!50, minimum size = 12mm] at (6.5, -2.65) {$J_2$};
	\node[circle, fill = red!50, minimum size = 14mm] at (9.3, -2.4) {$J_1$};

	\node[circle, fill = blue!50, minimum size = 14mm] at (0.7, -1.1) {$I_1$};
	\node[circle, fill = blue!50, minimum size = 12mm] at (3.5, -0.9) {$I_2$};
	\node[circle, fill = blue!50, minimum size = 9mm] at (6.5, -0.65) {$I_3$};
	\node[circle, fill = blue!50, minimum size = 6mm] at (9.25, -0.4) {};
	\node at (9.25, -0.4) {$I_4$};
	
	\node[circle, fill=black, inner sep = 0.6mm] (s) at (0.4, -0.7) {};
	\node (src) at (-0.2, 0.7) {\large$\source$};
	\draw (s) -- (src);

	\node[circle, fill=black, inner sep = 0.6mm] (t) at (9.6, -2.7) {};
	\node (trg) at (10.6, -4.1) {\large$\target$};
	\draw (t) -- (trg);

	\node[circle, draw, fill=black, inner sep = 0.6mm, label=left:{\large$c$}] at (3, -1.7) {};
	\node[circle, draw, fill=black, inner sep = 0.6mm, label=right:{\large$d$}] at (4, -2.3) {};

	\node[circle, draw, fill=black, inner sep = 0.6mm] (c) at (4.3, -1.4) {};
	\node at (4.05, -1.4) {\large$c'$};
	\draw[line width = 0.25mm, ->, decorate, decoration={coil, amplitude=1mm, segment length=1.5mm, pre length = 0.5mm, post length=1mm}] (c) -- (4.3, -0.5);

	\node[circle, draw, fill=black, inner sep = 0.6mm] (d) at (7.3, -2) {};
	\node at (7.05, -2) {\large$d'$};
	\draw[line width = 0.25mm, ->, decorate, decoration={coil, amplitude=1mm, segment length=1.5mm, pre length = 0.5mm, post length=1mm}] (7.3, -2.9) -- (d);
\end{tikzpicture}

%% file: chapters/conclusions.tex
We have looked at the reachability problem for
one-counter automata with equality and disequality tests.
We have proposed the idea of local inductive invariants and combined them
with the notion of unboundedness within an SCC.
Our construction circumvents the lack of computationally tractable descriptions:
indeed, in the subset sum example (Fig.~\ref{fig:subset-sum}~(left) and \cref{tractable}) the reachability set has exponential size, depending on $a_1, \ldots, a_n$.
There is no obvious means of compression available, and guessing/storing a traditional invariant is prohibitively expensive even for moderate~$n$.

An outstanding theoretical question is characterisation of complexity
of reachability in OCA with disequality tests. We have placed the problem
in \pitwop and, in the presence of equality tests, in \deltathreep. Both problems have already been known to be \NP-hard.
Are they \NP-complete or \coNP-hard too?
We also leave it open whether our technique can be extended to other systems
and settings, e.g., to parameter synthesis questions (see, e.g.,~\cite{GollerHOW10,LechnerMOPW18,PerezR22}).

In a more practical direction,
while the general invariant-based effective procedure for (non-)reachability
in vector addition systems~\cite{Leroux10} has not, to the best of our knowledge, been implemented,
our work identifies these potentially practial ways to reduce the search space for invariants in VASS.
The idea of restricting invariant sets to just a small `core' (in our case: a union of arithmetic progressions),
combined with the compositionality of invariants,
can help to direct an exploration of the search space, or assist a learning algorithm.

%% file: appendix.tex
\newpage 

\section{Details for Section \ref{sec:bg-oca}}
\label{app:bg-oca}
\input{appendix/background-details}

\section{Details for Section \ref{sec:pessimistic}}
\label{app:pessimistic}
\input{appendix/pessimistic-details}

\section{Details for Section \ref{sec:strongly-connected-OCA}}
\label{app:strongly-connected-OCA}
\input{appendix/strongly-connected-details}

\section{Details for Section \ref{sec:several-SCCs}}
\label{app:global-proofs}
\input{appendix/leaky-details}

%% file: appendix/background-details.tex
\subsection{Finding Positive-Effect Simple Cycles is \NP-hard}
\label{app:finding-simple-cycles-is-hard}

\begin{proposition} \label{finding-positive-simple-cycle-NPcomplete}
	Deciding, for a given OCA without tests $\Aa$ and a given state $q$, whether there exists a positive-effect simple $q$-cycle in $\Aa$ is an \NP-complete problem.
\end{proposition}
\begin{proof}
	Membership in \NP is obtained by using the $q$-cycle itself as a certificate.
	To prove \NP-hardness, we provide a reduction from the Hamiltonian path problem.
	Let $G = (V, E)$ be a directed graph and let $s, t \in V$ be two distinct vertices.
	A path from $s$ to $t$ is \emph{Hamiltonian} if it is simple and visits every vertex in the graph.
	The Hamiltonian path problem takes as input a directed graph $G = (V, E)$ and two vertices $s, t \in V$ and asks whether there is a Hamiltonian path from $s$ to $t$ in $G$.

	For the remainder of this proof, we fix an instance of this problem formed by $G = (V, E)$ and $s, t \in V$.
	Let $n = \size{V}$.
	We will now construct an OCA without tests (a 1-VASS) $\Aa = (Q, \transitions)$.
	Define $Q \coloneqq V \cup \set{q}$, where $q \not\in V$ is a new state, and 
	\begin{equation*}	
		\transitions \coloneqq \set{ (u, 1, v) : (u, v) \in E } \cup \set{ (q, 0, s), (t, -(n-2), q) }.
	\end{equation*}
	The construction of $\Aa$ takes polynomial time.
	We claim that there exists a Hamiltonian path from $s$ to $t$ in $G$ if and only if there exists a positive-effect simple $q$-cycle in $\Aa$.

	Suppose there exists a Hamiltonian path $\pi$ from $s$ to $t$ in $G$.
	Since $\pi$ visits every vertex in $G$, we have $\length{\pi} = n-1$.
	Consider the path $\sigma$ in $\Aa$ that is obtained from $\pi$ by replacing each edge $(u, v) \in E$ with the corresponding transition $(u, 1, v) \in \transitions$ as well as prepending the transition $(q, 0, s)$ and appending the transition $(t, -(n-2), q)$.
	Given that $\pi$ is a simple path in $G$, we know that $\sigma$ is a simple $q$-cycle in $\Aa$.
	Furthermore, given that $\length{\pi} = n-1$, we know that $\effect{\sigma} = 0 + n - 1 - (n-2) = 1$, so $\sigma$ has positive effect.

	Conversely, suppose there exists a positive-effect simple $q$-cycle $\sigma$ in $\Aa$.
	This $\sigma$ must begin with $(q, 0, s)$, the only outgoing transition from~$q$,
	and end with $(t, -(n-2), q)$, the only transition leading back to~$q$.
	Let $\sigma = (q, 0, s) \, \sigma' \, (t, -(n-2), q)$ for some $\sigma'$.
	Given that $\effect{\sigma} \geq 1$ and all other transitions in $\Aa$ have effect 1, we know that $\length{\sigma'} \geq n-1$.
	Since the cycle $\sigma$ is simple and $|Q \setminus \{q, s, t\}| = n-2$, we conclude that $\sigma'$ visits each of these $n-2$ states exactly once.
	So
	the path $\pi$ obtained from $\sigma'$ by replacing each transition $(u, 1, v) \in \transitions$ with the corresponding edge $(u, v) \in E$
	is a Hamiltonian path from $s$ to $t$ in~$G$.
\end{proof}

\subsection{Missing Proofs}

\sclifting*
\begin{proof}
	First, if $\source \step{*} \target$, then $\candidate{\source}{\target}$. 
	Conversely, let $\pi$ be a path such that $\candidaterun{\source}{\pi}{\target}$. 
	Let $(s,v) = \source$. 
	We claim that there is a positive cycle $\gamma_+$ such that, for all $k$, $(\gamma_+)^k$ yields a valid run from $\source$ (we say that $\gamma_+$ can be ``pumped up'' from $\source$). 
	Given that $\source$ is unbounded in $\A$, there exists a state $q \in Q$, a path $\pi$, and $z > v + \norm{\guards} + \norm{\transitions}\cdot|Q| + 1$ such that $\source \step{\pi} (q,z)$.
	Further, given that \A is strongly connected, there is a path $\sigma$, of length at most $|Q|$, from $q$ back to $s$; let $\gamma_+ \defeq \pi \cdot \sigma$. 
	We know that, because $z \geq \norm{\guards} + \norm{\transitions}\cdot |Q|$, following $\sigma$ from $(q,z)$ yields a valid run; suppose that run ends at the configuration $(s, v')$.
	Notice that $v' > v+\norm{\guards}$, so the cycle $\gamma_+$ from $(s,v')$ also yields a valid run, because all observed counter values are still greater than the disequality guards.
	Let $w_+>0$ be the effect of $\gamma_+$.  
	Symmetrically, there is a cycle $\gamma_-$, with effect $w_-<0$, that can be ``pumped down'' arbitrarily many times to $\target$.  
	By letting $m> \drop{\pi} + \norm{\guards}$, the path $(\gamma_+)^{m |w_-|} \cdot \pi \cdot (\gamma_-)^{m w_+}$ yields a run from $\source$ to $\target$. 
\end{proof}

\identify*
\begin{proof}
	The proof will use 1-dimensional vector addition systems with states (1-VASS).
	These are one-counter automata (as defined in this paper) without any tests.
	\begin{itemize}
	\item
	Syntactically, a 1-VASS is a pair $(Q, T)$, where $Q$ is the set of states and $T \subseteq Q\times \ZZ \times Q$ is the set of transitions.
	\item
	The semantics is the same as that of an OCA with disequality tests $(Q, T, \texttt{true})$, where the map $\texttt{true}$ assigns true to all states in $Q$.
	\end{itemize}
	We will consider an auxiliary problem which takes as input a 1-VASS $(Q, T)$, a state $q \in Q$, and a natural number $d$ (encoded in binary); the problem asks whether there is a positive-effect $q$-cycle $\gamma$ of length at most $\size{Q}$ such that $\drop{\gamma} \leq d$.
	We show that this problem can be solved in polynomial time, given that coverability in 1-VASS can be decided in polynomial time.
	Coverability is placed in~\Pcomplexity
	by a standard reduction from coverability to unboundedness (see, e.g.,~\cite[Lemma~1]{AlmagorCPS020})
	and a polynomial-time algorithm for unboundedness by Rosier and Yen~\cite[Theorem~3.4]{RosierY86}.
	Notice a stronger result from~\cite{AlmagorCPS020} is that coverability in 1-VASS is in fact in $\mathsf{NC^2} \subseteq \Pcomplexity$.

	\proofsubparagraph{Reduction of the auxiliary problem to coverability.}
	Let $n = \size{Q}$.
	We can construct an instance of coverability as follows.
	Consider the unfolding $(Q', T')$ where $Q' = \set{ q^{(i)} : q \in Q \text{ and } i \in [0, n] }$ and $T' = \set{ (p^{(i-1)}, a, q^{(i)}) : (p, a, q) \in T \text{ and } i \in[1, n]}$.
	Observe that there exists a positive-effect $q$-cycle whose length is at most $\size{Q}$ and with a drop bounded by $d$ in $(Q, T)$ if and only if $(q^{(i)}, d+1)$ can be covered from $(q^{(i)}, d)$ in $(Q', T')$ for some $i \in [1, n]$.
	Moreover, in that case, such a cycle is obtained directly from a path in the unfolded 1-VASS that witnesses coverability.

	\proofsubparagraph{Polynomial-time algorithms for minimum drop and membership in $\boldsymbol{\posconfigs}$.}
	We complete the proof of~\cref{lem:define-canonical}.
	\begin{itemize}
	\item
	First, observe that the minimum drop can be computed by a binary search for $d$.
	Let $m = \max\set{\size{a} : (p, a, q) \in T}$.
	By starting from an upper bound of $n(m+1)$, $d$ can be computed using a polynomial number (at most $\lceil \log(n(m+1)) \rceil$) of coverability queries.
	\item
	Second, to decide membership of a configuration $(q, v)$ in $\posconfigs$, it suffices to check that $q \in Q_+$, to compute $\drop{\gamma_q}$, and to check that $v \geq \drop{\gamma_q}$.
        \qedhere
	\end{itemize}
\end{proof}

\chaincount*
\begin{proof}
	As observed in~\cite[Remark 6]{AlmagorCPS020}, for 
	each $q\in Q$, the number of bounded $q$-chain is at most~$2|Q|$. Indeed, each guard seen in $\gamma_q$  this can be seen by pulling forwards the disequality guards of each state along the simple cycle~$\poscycle q$ to $q$. Each of these guards, say $x \neq g$, induces two bounded chains, namely the singleton~$\{(q,g)\}$ and the set of configurations below the guard with counter values congruent to $g$ modulo~$\effect{\poscycle q}$.
\end{proof}

%% file: appendix/pessimistic-details.tex
\subsection{Flows and Paths}

We start by formally defining the concept of flow.  
Given a function $f: \transitions \rightarrow \nats$ and $q \in  \states$, the \emph{flow entering $q$} is $\flowin{q}{f} := \sum_{\atrans = (q',a,q) \in \transitions} f(\atrans)$ and the \emph{flow leaving $q$} is $\flowout{q}{f} := \sum_{\atrans = (q,a,q') \in \transitions} f(\atrans)$. The \emph{underlying undirected graph} of $f$ is the graph $\graphof{f} := (Q',E)$ such that:
\begin{itemize}
    \item $Q'$ contains all states $q \in Q$ such that either $\flowin{q}{f} \ne 0$ or $\flowout{q}{f} \ne 0$, and
    \item for all $q,q' \in Q$, $\set{q,q'} \in E$ whenever there exists $\atrans \in \transitions$ such that $f(\atrans) >0$ and such that $\atrans = (q, a, q')$ or $\atrans = (q', a, q)$ for some $a \in \ZZ$. 
\end{itemize}
In words, $\graphof{f}$ has as edges the undirected and unlabelled version of transitions in $\A$ that have value greater than $0$ for $f$, and as vertices all the states that are incident to these transitions.
This graph is \emph{connected} if there is a path between any two vertices. 

\begin{definition}
\label{def:flow}
A \emph{flow} from $p \in \states$ to $q \in \states$ is a mapping $f: \transitions \rightarrow \nats$ such that:
\begin{enumerate}
    \item if $p = q$, then $\flowin{p}{f} = \flowout{p}{f}$;
    \item if $p \neq q$, then $\flowout{p}{f} = \flowin{p}{f} +1$ and $\flowin{q}{f} = \flowout{q}{f} + 1$;
    \item for every $q' \in \states\setminus\set{p, q}$, $\flowin{q'}{f} = \flowout{q'}{f}$; and
    \item $\graphof{f}$ is connected and contains $p$ and $q$.
\end{enumerate}
\end{definition}

Let $f$ be a flow from $p$ to $q$.
We may refer to $p$ as the \emph{start} of $f$ and $q$ as the end of $f$.
The \emph{size} of $f$ is $\size{f} \coloneqq \sum_{t \in \transitions} f(t)$.
Its \emph{effect} is $\effect{f} \coloneqq \sum_{\atrans \in \transitions} \effect{\atrans} \cdot f(\atrans)$.

A \emph{decomposition} of a flow $f$ from $p$ to $q$ is a finite sequence of flows $f_1, \dots, f_m$ such that $\sum_{i=1}^m f_i = f$, the start of $f_1$ is $p$, the end of $f_m$ is $q$, and, for all $i \in [1, m-1]$, the end of $f_i$ is the start of $f_{i+1}$.

We can compare two flows $f$ and $g$: we say that $f \leq g$ if $f(t) \leq g(t)$ for all $t \in \transitions$.
A~flow $f$ \emph{admits a positive cycle} if there exist $q \in \states$ and a flow $f_+ \leq f$ from $q$ to $q$ such that $\effect{f_+} > 0$.

Let $\apath$ be a path from $p$ to $q$.
We define the function $\flowof{\sigma}$ that assigns, to each transition $t$, the number of occurrences of $t$ in $\sigma$. 

\begin{lemma}
    Let $\apath$ be a path from $p$ to $q$; then $\flowof{\sigma}$ is a flow from $p$ to $q$.
\end{lemma}
\begin{proof}
    This follows immediately by induction on the prefixes of the path. 
\end{proof}

We call $\flowof{\apath}$ the \emph{flow of $\apath$}.
Similarly, the \emph{flow of a run} is the flow of the path underlying the run. 

\begin{lemma}
    \label{build_path_from_flow}
   For every flow $f$, there is a path $\apath$ such that $\flowof{\sigma} = f$.
\end{lemma}
\begin{proof}    
    We will prove, by induction, that for every $n \in \NN$, for every flow $f$ of size $n$, that there exists a path $\apath$ of length $n$ such that $\flowof{\apath} = f$.

    \proofsubparagraph*{Base case $n = 0$.}
    The only flow $f$ of size $0$ has $f(\atrans) = 0$ for all $\atrans \in \transitions$.
    Let $\apath$ be the empty path; $\flowof{\apath}[t] = 0$ for all $\atrans \in \transitions$.
    Hence $\flowof{\apath} = f$.

    \proofsubparagraph*{Inductive step $n > 0$.}
    Let $f$ be a flow from $p$ to $q$ of size $n$.
    For every $t \in \transitions$ such that $f(t) >0$, we denote by $g_t: \transitions \to \nats$ the function defined by $g(\atrans') = f(\atrans')$ for all $\atrans' \in \transitions \setminus \set{\atrans}$ and $g(\atrans) = f(\atrans)-1$.
    First, we will justify that there is $t = (q',a,q) \in \transitions$ such that $g_t$ is a flow from $p$ to $q'$. 
    There are two conditions to be a flow.
    These are the balance conditions (conditions 1, 2, and 3 of~\cref{def:flow}) and the connectivity condition (condition 4 of~\cref{def:flow}). 

    We first prove that, for every $t = (q', a, q)$, $g_t$ satisfies the balance conditions.
    It suffices to prove the conditions at $q$ and $q'$. 

    First, suppose $p = q$. 
    In this case, $\flowin{q}{f} = \flowout{q}{f}$ and $\flowin{q'}{f} = \flowout{q'}{f}$.
    By definition of $g_t$, $\flowin{q}{g_t} - 1 = \flowout{q}{g_t}$ and $\flowin{q'}{g_t} = \flowout{q'}{g_t} - 1$.
    Hence $g_t$ satisfies the balance conditions to be a flow from $p$ to $q'$. 
    Second, suppose $p \neq q$.
    In this case, $\flowin{q}{f} = \flowout{q}{f} + 1$, and $\flowin{q'}{f} = \flowout{q'}{f}$.
    Here, there is an additional case distinction: whether $p = q'$ or not.
    Suppose $p = q'$.
    In this case, we know that $\flowin{q}{g_t} = \flowout{q}{g_t}$.
    Furthermore, $\flowin{q'}{g_t} = \flowin{q'}{f} = \flowin{p}{f} = \flowout{p}{f} - 1 = \flowout{q'}{f} - 1 = \flowout{q'}{g_t}$. Because $q' = p$, this also means that $\flowin{p}{g_t} = \flowout{p}{g_t}$, so that $g_t$ satisfies the balance conditions to be a flow of size $n-1$ from $p$ to $q' = p$.
    Now, suppose $p \neq q'$.
    Since $g_t(t) = f(t) - 1$, we know that, again, $\flowin{q}{g_t} = \flowout{q}{g_t}$.
    However, this time, $\flowin{p}{g_t} + 1 = \flowout{p}{g_t}$ is unchanged but $\flowin{q'}{g_t} -1 = \flowout{q'}{g_t}$.
    Hence $g_t$ satisfies the balance conditions of a flow of size $n-1$ from $p$ to $q'$.

    We have proven that, for every $t$ leading to $q$, the function $g_t$ satisfies the balance conditions of a flow from $p$ to $q'$. 
    It remains to prove that there exists a transition $t$ such that $g_t$ satisfies the connectivity condition. 
    First, if there is a self-loop transition $t = (q, a, q)$ such that $f(t) > 0$, then $g_t$ immediately satisfies the connectivity condition.
    Now, suppose that there is no such self-loop at $q$. 
    Let $T_q$ denote the set of transitions of $\transitions$ that lead to $q$. 
    First, observe that $\transitions_q \neq \emptyset$.
    Since $\graphof{f}$ is connected, there is an edge of this graph that contains $q$. 
    Because $\flowout{q}{f} \leq \flowin{q}{f}$, there a transition $t \in \transitions_q$ such that $f(t) > 0$. 
    If $\transitions_q$ only contains one transition $t$, then $q$ only has one incident edge in $\graphof{f}$.
    The graph $\graphof{g_t}$ is the same as $\graphof{f}$ except potentially with $q$ removed if $f(t) = 1$ (otherwise, if $f(t) > 1$, $q$ remains in the graph). Therefore, $g_t$ satisfies the connectivity condition.
    
    Now, suppose that $\transitions_q$ contains at least two transitions.  
    Because $f$ is a flow, we have $\flowin{q}{f} \leq \flowout{q}{f}+1$.
    There exists $t \in \transitions_q$ such that $f(t) > 0$.
    For all $r \in Q$ such that $r \neq q$, $\flowin{r}{f} \leq \flowout{r}{f}$, so we can build a path by starting with $t$ and iteratively selecting new transitions until we arrive back at $q$. 
    This results in a $q$-cycle $\alpha$ such that $\flowof{\alpha} \leq f$.
    Let $t'$ denote the last transition of $\alpha$ and let $q' \in Q$ such that $t' = (q', a, q)$.
    Given that, for all transitions $s$ in $\alpha$, we know that $f(s) > 0$, the removal of the edge $\set{q', q}$ from $\graphof{f}$ does not change its connectivity.
    The graph $\graphof{g_{t'}}$ is either the same as $\graphof{f}$ or it has the edge $\set{q',q}$ removed (depending on whether $f(t') > 1$ or $f(t') = 1$).
    Regardless, $\graphof{g_{t'}}$ is connected.

    We have proven that there is $t = (q',a,q)$ such that $g_t$ is a flow from $p$ to $q'$ of size $n-1$.
    By the inductive hypothesis, there exists a path $\apath'$ of length $n-1$ from $p$ to $q'$.
    Therefore, the path $\apath = \apath'\,t$ is from $p$ to $q$, has length $n$, and satisfies $\flowof{\apath} = f$.
\end{proof}

\subsection{There are No Positive Cycles in the Flows of Pessimistic Runs}

We now prove that pessimistic runs may have no cycle in their flow. 
To do so, we need the following technical result inspired by \cite[Proposition 6]{HaaseKOW09}.

\begin{lemma}
    \label{cycle_of_drop_zero}
    Let $\gamma_+$ be a positive cycle of length $\ell$ ($\ell$ transitions, $\ell$ states) and, for all $i \in \nset{1}{\ell}$, let $\gamma_i$ be the cycle obtained when concatenating the suffix of $\gamma_+$ of length $\ell -i+1$ with the prefix of $\gamma_+$ of length $i-1$ ($\gamma_i$ corresponds to $\gamma_+$ but starting from and finishing at the $i$th state). Then there exists $k \in \nset{1}{\ell}$ such that $\drop{\gamma_k}= 0$. 
\end{lemma}
\begin{proof}
    For all $j \in \nset{1}{\ell}$, let $\gamma_{i,j}$ the path composed of the first $j$ transitions of $\gamma_i$. Let $k \in \nset{1}{\ell}$ such that $\effect{\gamma_{1,k}} = \min_i \effect{\gamma_{1,i}}$. We claim that $\drop{\gamma_k}= 0$, \emph{i.e.}, that, for all $j$, $\effect{\gamma_{k,j}} \geq 0$. 
    If $j\leq \ell-k+1$ then $\gamma_{1,k} \cdot \gamma_{k,j} = \gamma_{1,k+j}$ hence $\effect{\gamma_{1,k}} + \effect{\gamma_{k,j}} = \effect{\gamma_{1,k+j}}$ but $\effect {\gamma_{1,k+j}}\geq \effect{\gamma_{1,k}}$ so $\effect{\gamma_{k,j}} \geq 0$. If $j > \ell-k+1$, then $\gamma_{k,j} = \gamma_{k,\ell-k+1} \cdot \gamma_{1, j-\ell+k-1}$ so $\effect{\gamma_{k,j}} = \effect{\gamma_{k,\ell-k+1}} + \effect{\gamma_{1, j-\ell+k-1}} \geq \effect{\gamma_{k,\ell-k+1}} + \effect{\gamma_{1, k}} = \effect{\gamma_+} \geq 0$. We have proven that, for all $j$, $\effect{\gamma_{k,j}} \geq 0$ therefore $\drop{\gamma_k} \geq 0$.
\end{proof}

\subsection{Proof of Lemma \ref{certificate_pessimistic_runs_cc}}
\label{app:pessimistic-np}

\begin{proof}[Proof of~\cref{certificate_pessimistic_runs_cc}]
    Combine the following two lemmas (\cref{pessimistic_run_no_cycle_in_flow} and~\cref{no_positive_cycle_certificate}).
\end{proof}

\begin{lemma}
    \label{pessimistic_run_no_cycle_in_flow}
    The flow of a pessimistic run admits no positive cycle.
\end{lemma}
\begin{proof}
    Let $\run$ a pessimistic run, $f$ its flow and assume that $f$ admits a positive cycle; let $f_+ \leq f$ a flow from $q$ to $q$ such that $\effect{f_+} > 0$. We build a positive cycle $\pi_+$ on $q$ from $f_+$ thanks to \cref{build_path_from_flow}. By iteratively deleting negative simple cycles from $\pi_+$, we obtain a positive simple cycle $\gamma_+$. By \cref{cycle_of_drop_zero}, one can reorder the transitions of $\gamma_+$ to obtain a positive cycle $\gamma_+'$ on some state $q'$ such that $\drop{\gamma_+'} = 0$. This means that any visit to $q'$ is a visit to $\posconfigs$. Because $\gamma_+$ was built by deleting edges from $\gamma_+$, there is a transition in $f$ that ends on $q'$: $q'$ is visited in $\run$ after its first configuration which contradicts the fact that $\run$ is pessimistic. 
\end{proof}

For the next lemma, we recall that the size of an OCA~$\VASS$ with disequality tests, denoted~$\size{\VASS}$,
is defined as the bit size of its encoding, where all numbers are written in binary.

\begin{lemma}
    \label{no_positive_cycle_certificate}
    If the flow of a  run $\run: (q,x) \step{*} (q',y)$ admits no positive cycle, then $(q,x) \step{*} (q',y)$ admits a certificate verifiable in time polynomial in $\log(x+y+1) +\size{\VASS}$.
\end{lemma}
\begin{proof}
    Thanks to \cref{exponential_bound_reachability}, if such a $\run$ exists then we can assume that its length is bounded by $M:= \expbound(\size{\VASS}) +x + y$. Therefore, its flow $f$ satisfies $f \leq \expbound(\size{\VASS})+x+y$ hence is storable in space $\size{\transitions} \log(\expbound(\size{\VASS}+x+y))$. We assume that all flows described below are bounded by $\expbound(\size{\VASS})+x+y$.

    A \emph{certificate} for the existence of a run from $(q_s,x)$ to $(q_f,y)$ consists in two things. First, a flow $f$ from $q_s$ to $q_f$ of effect $y-x$ with no positive cycle and of largest value bounded by $\expbound(\size{\VASS}) + x +y$; additionally, a decomposition of $f$ into $f_1, \dots, f_m$ such that $m \leq 4  \size{\states}+1$. The decomposition must moreover satisfy some conditions stated below. Let $q_s = q_0, q_1, \dots, q_m = q_f$ the corresponding states, so that $f_i$ is a flow from $q_{i-1}$ to $q_i$. For all $i$, let $w_i := \sum_{j=0}^{i} \effect{f_i}$ and $x_i := w_i +x$. 
    For all $q \in \states$, let $I(q) := \set{i \in \nset{0}{m} \mid q_i = q}$ and $X(q) := \set{x_i \mid i \in I(q)}$.
    Also, for all $q$ such that $I(q) \ne \emptyset$, let $m_q := \min I(q)$ and $M_q := \max I(q)$. 
    First, we impose that, for all $i$, $x_i \geq 0$ and $x_i \notin \diseqtests(q_i)$ (all values $x_i$ are non-negative and do not hit disequality guards). 
    Moreover, for every state $q$ appearing in the flow: 
    \begin{enumerate}
    \item \label{item:iq_not_empty} $I(q) \ne \emptyset$; and
    \item \label{item:no_q_after} for all $i \geq M_q$, $q$ does not appear in $f_i$ except in one transition leaving $q$ for $i = M_q$; and
    \item \label{item:no_q_before}for all $i \leq m_q$, $q$ does not appear in $f_i$ except in one transition entering $q$ for $i = m_q$.
    \end{enumerate}
    The previous set of conditions guarantees that the counter value stays above $0$. 

    Additionally, for every state $q$ appearing in $f$ that has some disequality guard $g \geq 0$, if there exist $x,y \in X(q)$ such that $x < g < y$ (\emph{i.e.,}, the flow crosses $g$), then there exists $i,j \in I(q)$ such that $i<j$, $I(q) \cap \nset{i}{j} = \set{i,j}$ and: 
    \begin{enumerate}[resume]
    \item \label{item:g_crossed}$x_i > g > x_j$;
    \item \label{item:no_q_in_between} $q$ does not appear in flows $f_{i+1}$ to $f_{j}$, except in one transition leaving $q$ in $f_{i+1}$ and one transition entering $q$ in $f_j$. 
    \end{enumerate}
    These conditions guarantee that the disequality guards are not hit. Note, that if all values $x_i$ are above (resp. below) $g$, then the last (resp. first) occurrence of $q$ not hitting the guard witnesses that the guard is never hit.  

    We now prove that $(q_f,y)$ is reachable from $(q_s,x)$ if and only if there is a certificate for the existence of a run from $(q_s,x)$ to $(q_f,y)$.

    Suppose that there is a run $\run$ from $(q_s,x)$ to $(q_f,y)$. Let $f$ denote the flow of $\run$. In order to build the required decomposition of $f$, we simply split $\run$ into successives runs. First, we split $\run$ on every last first and occurrence of a state.
    Moreover, for every state $q$ that has a disequality guard $g$, we split run on the last visit to state $q$ above $g$, if it exists, and on the first visit to state $q$ below $g$, if it exists. Doing so, we have divided $\run$ into at most $4 \size{\states} +1$ consecutive runs $\run_1, \dots, \run_m$; we simply let $f_i$ the flow of $\run_i$ and all conditions above are trivially satisfied.

    Conversely, assume that we have a flow $f$ from $q_s$ to $q_f$ of effect $y-x$ and a decomposition $f_1 \dots f_m$ with $m \leq 4 \size{\states} +1$ that satisfies the conditions above. Let $q_s = q_0, q_1, \dots, q_m = q_f$ such that, for all $i$, $f_i$ is a flow from $q_{i-1}$ to $q_i$. For all $i$, let $w_i := \effect{\sum_{j=1}^{i} f_j}$, and $x_i := w_i + x$. For every $i$, let $\apath_i$ a path obtained from $f_i$ using \cref{build_path_from_flow}. By immediate induction, for all $i$, there is a candidate run $\run_i$ from $(q_{i-1},x_{i-1})$ to $(q_i, x_i)$, which forms a candidate run $\run$ from $(q_s,x)$ to $(q_f,y)$.
    We will now prove that all candidate runs are actual runs, \emph{i.e.}, that all values remain non-negative and that no disequality guard is hit.  

    First, all values remain positive thoughout $\run$. Indeed, let $q \in \states$. Because $f$ does not have negative cycles, if there is $q \in \states$ such that $(q,x)$ and $(q,y)$ are seen in $\run$ in this order, then $x \geq y$. 
    Let $k := M_q$. Because $\run_k$ ends on $(q,x_k)$ and $x_k \geq 0$, all values in $\run$ on state $q$ until $(q,x_k)$ have non-negative counter value. By \cref{item:no_q_after}, $(q,x_k)$ is the last visit of $\run$ to $q$ hence all counter values on $q$ are positive. 

    Moreover, no disequality guard is hit. Let $g$ be a disequality guard on some state $q$. If for all $x \in X(q)$, $x > g$, then in particular this is true for $x_{M_q}$ which proves that $g$ is not hit. Similarly, if for all $x \in X(q)$, $x < g$, then in particular this is true for $x_{m_q}$ which is, by \cref{item:no_q_before}, the highest counter value on state $q$ in $\run$, so $g$ is not hit. 
    Suppose now that there exists $x, y \in X(q)$ such that $x < g < y$.  
    Let $i,j \in I(q)$ satisfying the conditions above. Before the visit to $(q,x_i)$ and after the visit to $(q,x_j)$, we know that $g$ is not hit because $f$ has no positive cycle. Thanks to \cref{item:no_q_in_between}, we have that there is no visit to $q$ in $\run$ between the visit to $(q,x_i)$ and the one to $(q,x_j)$, proving that $g$ is not hit.  

    We finally argue that one can check in polynomial time whether a given flow along with a decomposition are a certificate; in particular, one can check in polynomial time whether a flow has no positive cycle.   
\end{proof}

%% file: appendix/strongly-connected-details.tex
\subsection{Proof of Lemma \ref{reachable_in_chain_reasonable}}

First, there are at most $2 \size{Q}^2$ bounded chains by \Cref{rem-number-bounded-chains} above. It suffices to prove that the set of reachable configurations within a chain is an arithmetic progression of upper limit less than $\norm{\tau} + (n+1) \norm{\transitions}$. Let $\achain$ a chain on state $q$, let $\gamma_+ := \poscycle{q}$ and $w:=\effect{\gamma_+}$. For every $x<y$ such that $(q,x), (q,y) \in \achain$, we have that there is $k \geq 0$ such that $(q,x) \step{(\gamma_+)^k} (q,y)$; therefore, $(q,x) \equiv (q,y) \pmod{w}$. Moreover, if there is $z$ such that $x < z < y$ and $z \equiv x \pmod{w}$ then there exists $\ell < k$ such that $(q,x) \step{(\gamma_+)^\ell} (q,z)$ hence $(q,z) \in \achain$. This proves that $\achain$ is an arithmetic progression; we now bound its largest configuration $(q,x_{\max})$.
Because $\gamma_+$ is a simple cycle, $w \leq (\size{\states}+1) \norm{\transitions}$. By maximality of $x_{\max}$, $\gamma_+$ cannot be taken from $(q,x_{\max})$ but $(q,x_{\max}) \in \posconfigs$ therefore $\gamma_+$ must hit a disequality guard when applied from $(q,x_{\max})$ hence $x_{\max} \leq (\size{\states} +1 ) \norm{\transitions} + \norm{\diseqtests} \leq 2 \size{Q} \cdot \norm{\transitions} \cdot \norm{\diseqtests}$.
\qed

\subsection{The Polynomial Hierarchy}
\label{app:def-ph}

This hierarchy of computational complexity classes
was first introduced by Meyer and Stockmeyer~\cite{MS72, Stockmeyer76}, see also Wrathall~\cite{Wrathall76}. Let $\MySigma_0^{\Pcomplexity}\eqdef \MyPi_0^{\Pcomplexity}\eqdef \Pcomplexity$, 
comprising  all polynomial-time decision problems.
For all $n \in \mathbb{N}$, let $\MySigma^{\Pcomplexity}_n$ be the set of all decision problems~$S$ such that there exists a predicate~$M$ decidable in polynomial time and a polynomial $P$ satisfying that, for every instance $x$, we have that $x \in S$ if and only if
\begin{align*}
	Q_1 \, y_1 \in \{0,1\}^{P(|x|)} \, Q_2 y_2 \, \in  \{0,1\}^{P(|x|)} \, \ldots \,  Q_n \, y_n \in & \{0,1\}^{P(|x|)} :\;\\
 &M(x,y_1,\ldots,y_n)=\text{\texttt{YES}},
\end{align*} 
where $Q_1 Q_2\ldots Q_n \in \{\exists,\forall\}^n$ is a (strictly) alternating sequence of quantifiers
$\exists$ and $\forall$ that starts with $\exists$. The definition of $\MyPi^{\Pcomplexity}_n$ is identical except that the sequence of quantifiers starts with $\forall$.
In particular,  $\MySigma^{\Pcomplexity}_1 = \compclass{NP}$ and $\MyPi^{\Pcomplexity}_1=\coNP$. An equivalent definition uses oracles, by letting $\MySigma^{\Pcomplexity}_{0} = \Pcomplexity$ and for all $n \geq 0$, $\MySigma^{\Pcomplexity}_{n+1} = \NP^{\MySigma^{\Pcomplexity}_n}$. Said otherwise, a problem is in $\MySigma^{\Pcomplexity}_{n+1}$ when it can be solved in non-deterministic polynomial time using a $\MySigma^{\Pcomplexity}_{n}$ oracle. Similarly, $\MyPi^{\Pcomplexity}_{n+1} = \compclass{coNP}^{\MyPi^{\Pcomplexity}_n}$.

The \emph{polynomial hierarchy} is  $\compclass {PH}\eqdef \bigcup_{n\in \mathbb{N}} \MySigma_n^{\Pcomplexity}$,  or equivalently $\compclass {PH}\eqdef \bigcup_{n\in \mathbb{N}} \MyPi_n^{\Pcomplexity}$. In words, $\PH$ is the class of all  problems that can be decided with a constant number of alternating quantifiers.

By applying the quantifier-based definition to \sigmatwop, we obtain that a decision problem $S$ is in \sigmatwop such that there exists a predicate~$M$ decidable in polynomial time and a polynomial $P$ satisfying that, for every instance $x$, we have that $x \in S$ if and only if
\begin{align*}
	\exists \, y \in \{0,1\}^{P(|x|)}  \, \forall z \, \in  \{0,1\}^{P(|x|)} \, M(x,y,z)=\text{\texttt{YES}}.
\end{align*}

The oracle-based definition is that the class $\sigmatwop$ contains all decision problems that can be solved in non-deterministic polynomial time with access to an \compclass{NP} oracle. Note that, for our purposes, an \compclass{NP} oracle is also a \compclass{coNP} oracle, as it may also solve \compclass{coNP} problems.

\subsection{Technical Lemmas for Bounding Counter Values}
\label{app:exp_bounds}

\begin{lemma}
\label{exponential_bound_reachability}
For every OCA with equality and disequality tests $\VASS$ and every two configurations $\source = (s,x)$ and $\target = (t,y)$,
if $\reach{\source}{\target}$ then $\RUN{\source}{\pi}{\target}$ for some valid run~$\rho$ whose length and highest counter value are at most $\mathrm{exp}(\size{\VASS} + \log(x+1) + \log(y+1))$.
\end{lemma}

\begin{proof}
We transform the given $\VASS$ into another OCA, where updates are encoded in unary.
Each transition are replaced by a simple path of length equal to the absolute value of its update value.
An equality test $x = k$ can be first replaced with a sequence of three operations: decrement by $k$, equality test for~$0$,
and increment by $k$.
A disequality test $x \ne k$ can be directly converted into a sequence of $k+1$~decrements and $k+1$~increments,
connected by a simple gadget comprising of $k$ zero-test transitions and one ordinary transition.
The result is a new OCA without disequality tests (but possibly with zero tests), which mimics the behaviour of the original one
and in which all updates are encoded in unary.

We can also attach simple paths of length $x$ and $y$ to the obtained OCA so that the reachability query is for two configurations with counter value~$0$.
The size of the new OCA is bounded from above by $n \defeq \mathrm{exp}(\size{\VASS} + \log(x+1) + \log(y+1))$.

It is known that in every OCA with $n$ states and unary updates (as well as possibly with zero tests)
if there exists a run from the source configuration to the target configuration (with counter values~$0$),
then there also exists such a run of length $O(n^2)$ (see, e.g.,~\cite{ChistikovCHPW19}).
This run is mimicking a valid run in the original OCA with equality and disequality tests, so
the upper bound on the length (and thus on the counter value) carries over.
\end{proof}

\begin{lemma}
\label{pessimistic_run_bounds}
Suppose all configurations of a set $S \subseteq \configs$ have counter value bounded by $N \in \nats$. Then all configurations in $\poststarpessimistic{S}$ have counter value bounded by $N + (\size{\states}-1) \norm{\transitions}$.
\end{lemma}

\begin{proof}
Let $\run : (q,x) \step{*} (q',y)$ be a pessimistic run, with $(q,x) \in S$. 
We prove, by induction on the number $k$ of states visited in $\run$, that the largest counter value in $\run$ is at most $N + (k-1) \norm{\transitions}$. For $k=1$, because $(q,x) \in S$, we have $x \leq N$.
Assume that $\run$ visits $k+1$ states; let $q_1, \dots, q_{k+1}$ the states appearing in $\run$ ordered by first visit in $\run$. We decompose $\run$ on its first visit to $q_{k+1}$: $\run: (q,x) \step{\apath_p} (q_p,x_p) \step{\atrans} (q_{k+1},x) \step{\apath_s} (q',y)$. 
Let $\run_p: \run: (q,x) \step{\apath_p} (q_p,x_p)$ and $\run_s: (q_{k+1},x) \step{\apath_s} (q',y)$. By induction hypothesis, all counter values in $\run_p$ are less than $N + (k-1) \norm{\transitions}$ therefore $x \leq N + k \norm{\transitions}$. We now prove it for $\run_s$. Let $(q_s, x_s)$ a configuration seen in $\run_s$. If $q_s = q_i$ for $i \leq k$, then $x_s \leq x_i \leq N + (k-1) \norm{\transitions}$ as it otherwise would imply a positive cycle in the flow of $\run$, contradicting  \cref{pessimistic_run_no_cycle_in_flow}. For the same reason, if $q_s = q_{k+1}$ then $x_s \leq x \leq N + k \norm{\transitions}$. The proves that all configurations in $\run$ are bounded by $N + k \norm{\transitions}$, concluding the induction step. 
\end{proof}

In particular, when checking the third condition of the invariant in \cref{subsec:strongly-connected-complexity}, one already know that $c \in I$ can be considered to be at most exponential because the counter values $I$ are bounded by an exponential in $\size{\VASS}$. \cref{pessimistic_run_bounds} therefore yields a bound on the counter value of the configuration $d$ to consider, because there must be a pessimistic run from $c$ to $d$. The same argument applies to the proof of \cref{thm:complexity}.

%% file: appendix/leaky-details.tex
\subsection{Proofs of Section \ref{sec:locally-bounded}}
\label{app:locally-bounded-proofs}

\lifting*
\begin{proof}
	First, if the instance is positive then there is a run and therefore a candidate run.
	In the other direction, let $\apath$ be a path corresponding to a candidate run from $c$ to~$d$. 
	Recall that the configuration $c$ is locally unbounded.
	Almagor~et~al.~\cite[Prop.~4]{AlmagorCPS020} prove that a configuration
	in an OCA with disequality tests is unbounded if and only if it can reach
	an unbounded configuration in a set denoted $\posconfigs$.
	(The set $\posconfigs$ in~\cite{AlmagorCPS020} is chosen based on a slightly different set of cycles
	 compared to the present paper, but the difference does not matter to us here.)
	By this result, there exists an infinite lasso-shaped run starting in~$c$, that is,
	whose path is of the form
	\begin{equation*}
	\pi \, \lambda^\omega \defeq
	\pi \, \lambda \,\lambda \,\lambda \ldots\,,
	\end{equation*}
	where $\pi$ and $\lambda$ are finite sequences, and moreover the latter corresponds to some positive cycle (call it $\gamma_+$)
	within the SCC of $c$.
	It is no loss of generality to assume that $\lambda$ has zero drop and starts in state~$q$.
	(If this is not the case, the cycle can be `rotated' appropriately; see, e.g., \Cref{cycle_of_drop_zero} in Appendix~\ref{app:pessimistic}.)
	Thus, from $c$ it is possible to reach some configuration $c'$, pump the cycle $\gamma_+$ many times and get back to $s$ using a path of less that $\size{\states}$ transitions. (Recall that $\pi$ and $\lambda$ stay in the SCC of $c$.)
	Therefore, we can build a positive $s$-cycle $\alpha_+$ that can be pumped to infinity from $c$.

	With a similar reasoning, we obtain in $\rev{\VASS}$ a positive $t$-cycle $\alpha_-$ that can be pumped down from infinity to $d$. Let $\apath$ be a candidate run from $c$ to $d$. Let $M \defeq \drop{\apath} + \norm{\diseqtests}$. We thus obtain a run from $c$ to $d$ by pumping the cycle $\alpha_+$ up $M \cdot \effect{\alpha_-}$ times, taking $\apath$ and pumping the cycle $\alpha_-$ down $M \cdot \effect{\alpha_+}$ times.
\end{proof}

\lpostnp*
\begin{proof}
	We want to provide a certificate the existence of a pessimistic and locally bounded run $c \xrightarrow[L]{\pi} d$. First, thanks to \cref{pessimistic_run_bounds}, counter values in such a run are bounded by $(\size{Q}-1) \norm{\transitions} + \size{c}$. We are therefore able to store a given configuration observed by the run in space polynomial in the size of the system and the size of $c$. 
	The certificate for $c \xrightarrow[L]{\pi} d$ consists in the following:
	\begin{itemize}
		\item the list $S_1, \dots, S_m$ of SCCs visited by $\pi$ in this order;
		\item for every SCC $S_i$, configurations $e_i$ and $e'_i$ where $e_i$ is the first configuration of the run in $S_i$ and $e'_i$ is the last configuration of the run in $S_i$,
		\item for every $i < m$, the transition $t \in \transitions$ from $S_i$ to $S_{i+1}$ that witnesses $e'_i \xrightarrow{t} e_{i+1}$;
		\item within a given SCC $S_i$, the certificate for the pessimistic run from $e_i$ to $e'_i$ as in \Cref{certificate_pessimistic_runs_cc}.
	\end{itemize}
	To verify such a certification in polynomial time, we verify that the previous elements indeed define a run $c \xrightarrow{*} d$; moreover, we verify that all configurations $e_i$ are locally bounded, which can be done in polynomial time thanks to \cref{decide_boundedness}.
\end{proof}

\subsection{Proof of Lemma \ref{source_target_simplification}}

\begin{proof}[Proof of~\cref{source_target_simplification}]
	The assumption that $\source \in \posconfigs \cap L$ and $\target \in \rev{\posconfigs} \cap \rev{L}$ follows from~\cref{clm:posconfigs-assumption}.
\end{proof}

\begin{numclaim}\label{clm:posconfigs-assumption}
	Let $\Aa$ be an OCA with disequality tests, and let $(p, x)$ and $(q, y)$ be two configurations.
	Then, there exists an OCA with disequality tests $\Bb$ and two configurations $(p', x)$ and $(q', y)$ such that:
	\begin{enumerate}
		\item
		$(p, x)\xrightarrow{*} (q, y)$ in $\Aa$ if and only if $(p', x) \xrightarrow{*} (q', y)$ in $\Bb$; and
		\item 
		$(p', x) \in \posconfigs \cap L$ and $(q', y) \in \rev{\posconfigs} \cap \rev{L}$.
	\end{enumerate}
\end{numclaim}
\begin{proof}
	Given an OCA with disequality tests $\Aa$ and two configurations $(p, x)$ and $(q, y)$, we will construct an OCA with disequality tests $\Bb$ by adding two states $p'$ and $q'$, and four transitions $(p', 0, p)$; $\ell_1 = (p', 1, p')$; $(q, 0, q')$; and $\ell_2 = (q', -1, q')$.
	We also add disequality tests $\neq x+1$ and $\neq y+1$ at $p'$ and $q'$, respectively.

	First, observe that, in $\Bb$, indeed $(p', x) \in \posconfigs$ since there is a positive-effect (single-transition) cycle $(\ell_1)$ with zero drop at $p'$ (and so it can be taken from any counter value, including $x$).
	Also, $(q', y) \in \rev{\posconfigs}$ since, in reverse, the single-transition cycle $(\ell_2)$ has positive effect and zero drop at $q'$ (and so it can taken from any counter value, including $y$).

	Second, it is also true that $(p', x) \in L$ in $\Bb$.
	This is true because $p'$ belongs to its own SCC, and even though there is the positive-effect transition $\ell_1$, we know that $(p', x)$ is locally bounded because of the disequality test added at $p'$ that forbids the counter from being equal to $x+1$.
	The same can be said for $(q', y) \in \rev{L}$ in $\rev{\Bb}$.

	If $(p, x)\xrightarrow{*}(q, y)$ in $\Aa$, then trivially $(p', x)\xrightarrow{*}(q', y)$ in $\Bb$ by appending and prepending the zero-effect transitions from $p'$ to $p$ and $q$ to $q'$ at the start and end of the run, respectively.

	Conversely, suppose that $(p', x)\xrightarrow{\pi}(q',y)$ in $\Bb$, for some path $\pi$.
	It is not possible for $\pi$ to use the transition $\ell_1$, otherwise the disequality test at $p'$ is violated.
	Also, $\pi$ does not use the transtion $\ell_2$.
	Suppose, for a moment, that $\ell_2$ is used.
	Since $(\ell_2)$ is the only cycle at $q'$ and it has negative effect, then it must be the case that the run $(p', x)\xrightarrow{\pi}(q',y)$ first observes a configuration $(q', y')$ for some $y' > y$.
	However, the only way to reach $(q', y)$ from $(q', y')$ is to use $(\ell_2)$.
	This means that the configuration observed prior to reaching $(q', y)$ would have to be $(q', y+1)$ which, however, violates the disequality tests at $q'$.
	Now, given that $(p', x)\xrightarrow{\pi}(q',y)$ does not use transitions $\ell_1$ and $\ell_2$, one can trim the zero-effect transitions from $p'$ to $p$ and $q$ to $q'$ away from the start and end of the run, respectively.
	This yields a run witnessing $(p,x)\xrightarrow{*}(q,y)$ in $\Aa$.
\end{proof}

\subsection{Discussion of the Choice of Operators}
\label{app:discussion-operators}

In \cref{con:inductive}, we made the choice of using operator $\bpposts{\cdot}$, and not $\pposts{\cdot} \cup \Post{\pposts{\cdot}}$ as in \cref{sec:strongly-connected-OCA}. 
Indeed, if we had used $\pposts{\cdot} \cup \Post{\pposts{\cdot}}$ instead, then in order to obtain completeness (\cref{lem:completeness}), one would have to change the perfect core. 
We want the perfect core to be contained in $L \cap \posconfigs$ so that it has a short representation; the natural candidate would be to take $\posts^*(\source) \cap L \cap \posconfigs$ (and symmetrically in $\rev{\A})$. 
This perfect core would satisfy the inductive property; however, this choice would break soundness (\cref{lem:soundness}). 
Indeed, this invariant could contain a locally bounded configuration $c$ that is reached from $\source$ using a run that visits many locally unbounded configurations in $\configs\setminus \posconfigs$ before coming back in $L$. 
In this case, it could be that $c$ is not captured by the inductive property, so one could find an inductive invariant $I$ that does not contain $c$.

\subsection{Proofs of Section \ref{sec:leaky:sound-and-complete}}
\label{app:sound-complete}

\soundness*
\begin{proof}
	Suppose $(I, J)$ is an inductive core. 	
	We know therefore that 
	\begin{equation*}
		\Post{\bpposts{I}} \cap \posconfigs \cap L \sset I.
		\end{equation*}
	Here, we will show that $B \cap \posconfigs \sset I$, showing $\rev{B} \cap \rev{\posconfigs} \sset J$ follows in the same way.

	For sake of contradiction, suppose there is a configuration in $B \cap \posconfigs$ that does not belong to $I$.
	In fact, among all such configurations, we will have a configuration $c$ that is witnessed by a locally bounded run $\source \xrightarrow[L]{\pi} c$ observing the fewest configurations in $\posconfigs$. 
	Note that since $\source \in I$, we may assume that $\length{\pi} \geq 1$.
	Suppose $\ell$ is the last configuration in the run that belongs to $I$ and suppose $p$ is the penultimate configuration in the run.
	Allow us to decompose the run into parts:
	\begin{equation*}
		\tempinit 
		\xrightarrow[L]{\theta} \ell
		\xrightarrow[L]{\sigma} p 
		\xrightarrow[L]{t} c,
	\end{equation*}
	where $\theta \,\sigma\,t = \pi$.

	We will now show that $\ell \xrightarrow[L]{\sigma} p$ is a pessimistic run.
	Indeed, if not, then this run observes a configuration $d \in \posconfigs$ such that $d \neq p$.
	Suppose 
	\begin{equation*}
		\ell \xrightarrow[L]{\sigma_1} d \xrightarrow[L]{\sigma_2} p, 
	\end{equation*}
	where $\sigma = \sigma_1\,\sigma_2$.
	Then since run $\source \xrightarrow[L]{\theta \,\sigma_2} d$ is locally bounded and by assumption $\ell$ was the last configuration in $I$, we know that $d \in (B \cap \posconfigs) \setminus I$.
	The run $\source \xrightarrow[L]{\theta \,\sigma_2} d$ observes fewer configurations in $\posconfigs$ than $\source \xrightarrow[L]{\pi} d$, contradicting its minimality.
	Therefore, $\ell \xrightarrow[L]{\sigma} p$ is a pessimistic run for it does not observe any configurations in $\posconfigs$.

	Now, since $\ell \in I$ and $\ell \xrightarrow[L]{\sigma} p$ is a pessimistic run, we know that $p \in \bpposts{I}$. 
	Hence, since $p \xrightarrow[L]{t} c$, it is true that $c \in \Post{\pposts{I}}$. 
	And so, since $B \sset L$, we know that $c \in \Post{\bpposts{I}} \cap L \cap \posconfigs \sset I$, contradicting the assumption that $c \notin I$.

	Therefore, for inductive cores $(I, J)$, $B \cap \posconfigs \sset I$ and $\rev{B} \cap \rev{\posconfigs} \sset J$. 
\end{proof}

\completeness*
\begin{proof}
	Let $I = B \cap \posconfigs$ and $J = \rev{B} \cap \rev{\posconfigs}$.
	Here, we will indeed show that $\Post{\bpposts{I}} \cap \posconfigs \cap L \sset I$, as per Condition~\ref{con:inductive}. 
	Showing that $\Pre{\bopres{J}} \cap \rev{\posconfigs} \cap \rev{L} \sset J$ follows in the same way.

	Suppose, for sake of contradiction, that $\Post{\bpposts{I}} \cap \posconfigs \cap L \not\sset I$; therefore
	\begin{equation*}
		\Post{\bpposts{ B \cap \posconfigs }} \cap \posconfigs \cap L \not\sset B \cap \posconfigs.
	\end{equation*}
	Thus, there exists a configuration $c$ belonging to $\Post{\bpposts{ B \cap \posconfigs }}$, $\posconfigs$, and $L$ such that $c$ either does not belong to $B$ or $\posconfigs$. 
	Clearly, since $c \in \posconfigs$, we may assume that $c \notin B$.
	Recall that $B$ is the set of configurations that can be reached from $\tempinit$ by a locally bounded run.

	Now, if $c \in \Post{\bpposts{B \cap \posconfigs}}$, then there exists $d \in B \cap \posconfigs$ and $p \in \bpposts{(p, x)}$ such that 
	\begin{equation*}
		\tempinit
		\xrightarrow[L]{\theta} d
		\xrightarrow[L]{\sigma} p
		\xrightarrow{t} c.
	\end{equation*}
	Where $\tempinit \xrightarrow[L]{\theta} d$ is the locally bounded run for $d \in B$; $d \xrightarrow[L]{\sigma} p$ is the locally bounded (and pessimistic) run for $p \in \bpposts{d}$; and there is the single transition run $p \xrightarrow{t} c$ for $c \in \Post{\bpposts{B \cap \posconfigs}}$.
	Notice that the final single transition run is also locally bounded because there is a locally bounded run to $p$ and $c \in L$ is assumed.
	Altogether, we observe that there is a locally bounded from from $\tempinit$ to $c$:
	\begin{equation*}
		\tempinit \xrightarrow[L]{\theta\,\sigma\,t} c.
	\end{equation*}
	Hence, $c \in B$, contradicting the opening assumption.
	We therefore conclude that \linebreak $\Post{\bpposts{I}} \cap \posconfigs \cap L \sset I$.
\end{proof}

\subsection{Proofs of Section \ref{sec:wrapping-up}}
\label{app:wrapping-up-proofs}

\leakyinvariant*

Before we prove~\cref{lem:leaky-invariant}, we first provide an auxiliary statement that will be used a few times in the proof.
In short, the following claim states (for the forwards direction) that all configurations that are reachable, from the initial configuration, via locally bounded runs are captured by $\pposts{I}$.

\begin{numclaim}\label{clm:run-in-i}
	Let $I$ and $J$ be sets of configurations such that $(I, J)$ is inductive (\cref{con:inductive}) and let $c$ be a configuration.
	\begin{enumerate}[label = (\arabic*)]
		\item If $\source\xrightarrow[L]{*}c$, then $c \in \pposts{I}$.
		\item If $c\xrightarrow[\rev{L}]{*}\target$, then $c \in \opres{J}$.
	\end{enumerate}
\end{numclaim}
\begin{proof}
	We will prove statement (1) and statement (2) follows symmetrically.
	Suppose $\source\xrightarrow[L]{*}c$.

	First, suppose $c \in \posconfigs$.
	Since $\source\xrightarrow[L]{*}c$ is a locally bounded run, we know that $c \in B$.
	Since $c \in B \cap \posconfigs$ and give that $(I, J)$ is inductive, by~\cref{lem:soundness} (soundness), we conclude that $c \in I \sset \pposts{I}$.

	Second, suppose $c \notin \posconfigs$.
	Let $\ell$ be the last configuration in $\source\xrightarrow[L]{*}c$ that belongs to $\posconfigs$.
	Such a configuration $\ell$ exists because $\source \in \posconfigs$ by~\cref{source_target_simplification}.
	Suppose $\source \xrightarrow[L]{\pi_1} \ell \xrightarrow[L]{\pi_2}$, where $\pi_1\,\pi_2 \pi$.
	Since $\source\xrightarrow[L]{\pi_1} \ell$, we know that $\ell \in B$.
	Thus, $\ell \in B \cap \posconfigs$, so given that $(I, J)$ is inductive, by~\cref{lem:soundness} (soundness), we find that $\ell \in I$.
	Furthermore, by choice of $\ell$, we know that no configurations in teh run $\ell \xrightarrow[L]{\pi_2} c$, except $\ell$, belong to $\posconfigs$.
	This means that $\ell \xrightarrow[L]{\pi_2} c$ is a pessimistic run, so $c \in \pposts{\ell} \sset \pposts{I}$, as required.
\end{proof}

\begin{proof}[Proof of~\Cref{lem:leaky-invariant}]
	\hfill

	\proofsubparagraph*{First direction.} If $\source$ cannot reach $\target$ in $\Aa$, one can construct a non-reachability witness $(I, J)$ for $\source$ and $\target$.

	Suppose, indeed, that $\source$ cannot reach $\target$ in $\Aa$.
	Let $I = B \cap \posconfigs$ and $J = \rev{B} \cap \rev{\posconfigs}$.
	Recall that $B$ is the set of all configurations that can be reached from $\source$ via locally bounded runs in $\Aa$ and $\rev{B}$ is the set of all configurations that can be reached from $\target$ via locally bounded runs in $\rev{\Aa}$.
	By~\Cref{lem:completeness} (completeness), $(I, J)$ is inductive, so the pair satisfies Condition~\ref{con:inductive}.

	Now, consider the induced $\finv  = \pposts{I} \cup \Post{\pposts{I}}$ and $\binv = \opres{J} \cup \Pre{\opres{J}}$.
	We will prove that $(\finv, \binv)$ form a separator, so the pair satisfies Condition~\ref{con:separator}.
	First, we will argue that all configurations in $\finv$ can be reached from $\source$.
	Consider $c \in \finv$, there must exist $d \in I$ such that $c \in \pposts{d}$ or $c \in \Post{\pposts{I}}$.
	Regardless of whether $c \in \pposts{I}$ or $c \in \Post{\pposts{I}}$, there is a run from $d$ to $c$.
	Since $d \in I \sset B$, there is a locally bounded run from $\source$ to $d$. 
	So, combining the two runs, one observes that $\source \xrightarrow{*} c$ in $\Aa$.
	The same argument can be used to show that all configurations in $c \in \binv$ can be reached from $\target$ in $\rev{\Aa}$; that means that $c \xrightarrow{*} \target$ in $\Aa$.

	Suppose that $(\finv, \binv)$ does not satify Condition~\ref{con:separator} \ref{sep1}.
	There exists $c \in \finv$ and $d \in \binv$ such that $c \xrightarrow{} d$.
	By the above, we know that $(s,x) \xrightarrow{*} c$ and $d \xrightarrow{*} (t,y)$; 
	\begin{equation*}
		(s,x) \xrightarrow{*} c \xrightarrow{} d \xrightarrow{*} (t, y)
		\; \implies \;
		(s,x) \xrightarrow{*} (t,y).
	\end{equation*}
	This creates a contradiction, so $(\finv, \binv)$ must satisfy Condition~\ref{con:separator} \ref{sep1}.

	Now, suppose that $(\finv, \binv)$ does not satify Condition~\ref{con:separator} \ref{sep2}.
	In this case, there exists $c \in \finv$ and $d \in \binv$ such that $c$ is locally unbounded in $\Aa$, $d$ is locally unbounded in $\rev{\Aa}$, and $\candidate{c}{d}$.
	By~\Cref{unbounded_rev_unbounded_global}, we know that actually $c \xrightarrow{*} d$ holds.
	Again, by the above, we know that $(s,x) \xrightarrow{*} c$ and $d \xrightarrow{*} (t,y)$; 
	\begin{equation*}
		(s,x) \xrightarrow{*} c \xrightarrow{*} d \xrightarrow{*} (t, y)
		\; \implies \;
		(s,x) \xrightarrow{*} (t,y).
	\end{equation*}
	This creates a contradiction, so $(\finv, \binv)$ must satisfy Condition~\ref{con:separator} \ref{sep2}.

	We now prove that $I$ and $J$ defined above have a concise description. 
	The argument is very similar to the one for the proof of \cref{unreachability_invariant_better}. 
	Indeed, $B$ cannot intersect unbounded chains, as all configurations in unbounded chains are locally unbounded. 
	Let $C$ be one of at most $2\size{Q}^2$ bounded chains.
	The intersection of $B$ with $C$ is an arithmetic progression whose largest value is bounded by an exponential in the size of the system. 
	Here, we can use~\cref{reachable_in_chain_reasonable}.

	\proofsubparagraph*{Second direction.} If there is a non-reachability witness $(I, J)$ for $\source$ and $\target$, then $\source$ cannot reach $\target$ in $\Aa$.

	Suppose that there exist sets of configurations $I$ and $J$ such that $\source \in I$; $\target \in J$; $(I,J)$ is inductive (Condition~\ref{con:inductive}); and the induced $\finv = \pposts{I} \cup \Post{\pposts{I}}$ and $\binv = \opres{J} \cup \Pre{\opres{J}}$ form a separator $(\finv, \binv)$ (Condition~\ref{con:separator}).
	We will argue that $\source$ cannot reach $\target$.

	Assume, for sake of contradiction, that $\source$ can reach $\target$ in $\Aa$, so there is a run $\RUN{\source}{\pi}{\target}$.
	We will decompose the run into parts about reahc of the $k$ SSCs that were visited.
	Precisely, we let $e_1, \ldots, e_k$ be the first, and $f_1, \ldots, f_k$ be the last, configurations observed in each SCC, respectively.
	\begin{equation*}
		\source = e_1 \xrightarrow{\pi_1} f_1 
		\xrightarrow{t_1} e_2 \xrightarrow{\pi_1} f_2 
		\xrightarrow{t_2} \cdots
		\xrightarrow{t_{k-1}} e_k \xrightarrow{\pi_k} f_k = \target
	\end{equation*}
	Note that $e_1, \ldots, e_k$ are the first, and $f_1, \ldots, f_k$ are the last, configurations observed in each SCC, respectively. 

	Consider the case where all $e_1, \ldots, e_k$ are locally bounded.
	This means that the entire run is locally bounded $\source \xrightarrow[L]{\pi} \target$.
	Since $\source \neq \target$, it is true that $\length{\pi} \geq 1$, so we can decompose the run $\source \xrightarrow[L]{\pi'} p \xrightarrow[L]{t} \target$, where $p$ is the penultimate configuration in the run.
	By~\cref{clm:run-in-i}, we know that $p \in \pposts{I} \sset \finv$.
	We know that $\target \in J$, and since $J \sset \binv$, we have $\target \in \binv$.
	This is a contradiction because $p \rightarrow \target$ bolds, but $(\finv, \binv)$ was assumed to satisfy Condition~\ref{con:separator}~\ref{sep1}.

	In is therefore the case that not all $e_1, \ldots, e_k$ are locally bounded.
	In fact, the same argument can be used to show that not all $f_1, \ldots, f_k$ are locally bounded in $\rev{\Aa}$. 
	Let $i \in \set{1, \ldots, k}$ be the first index such that $e_i$ is not a locally bounded configuration in $\Aa$ and let $j \in \set{1, \ldots, k}$ be the last index such that $f_j$ is not a locally bounded configuration in $\rev{\Aa}$.
	We will consider two subcases: when $i > j$ and when $i \leq j$.

	Before splitting into these the two subcases, given that $e_1 = \source$ and 
	\begin{equation*}
		e_1 \xrightarrow[L]{\pi_1\,t_1\,\cdots\,t_{i-2}\,\pi_{i-1}} f_{i-1},
	\end{equation*} 
	we know that $f_{i-1} \in \pposts{I}$ by~\cref{clm:run-in-i}.
	Furthermore, since $f_{i-1} \xrightarrow{t_{i-1}} e_i$, we deduce that $e_i \in \Post{\pposts{I}}$.
	Together, and to summarise, we know that both $f_{i-1}$ and $e_i$ belong to $\finv$.
	Likewise, given that $f_k = \target$ and again by~\cref{clm:run-in-i}, we can conclude that $f_j$ and $e_{j+1}$ belong to $\binv$.

	\emph{Case $i > j$.}
	We know that $f_j$ occurs before $f_{i-1}$ ($j \leq i-1$) so there is a locally bounded run from $\source$ to $f_j$:
	\begin{equation*}
		e_1 \xrightarrow[L]{\pi_1\,t_1\,\cdots\,t_{j-1}\,\pi_j} f_j \xrightarrow[L]{t_j \, \pi_{j+1} \, \cdots \, t_{i-2} \, \pi_{i-1} f_{i-1}}.
	\end{equation*}
	By~\cref{clm:run-in-i}, we conclude that $f_{i-1} \in \pposts{I} \sset \finv$.
	Above, we justified that $e_{j+1} \in \binv$.
	However, we know that $f_j \xrightarrow{t_j} e_{j+1}$.
	This creates a contradiction as $(\finv, \binv)$ was assumed to be a separator (Condition~\ref{con:separator}~\cref{sep1}).

	\emph{Case $i \leq j$.}
	Above, we justified that $e_i \in \finv$ and $f_j \in \binv$.
	Furthermore, by assumption $e_i$ is not locally bounded in $\Aa$, so $e_i \notin L$ and $f_j$ is not locally bounded in $\rev{\Aa}$, so $f_j \notin \rev{L}$.
	However, there is a run from $e_i$ to $f_j$:
	\begin{equation*}
		e_i \xrightarrow[L]{\pi_i \, t_i \, \cdots t_{j-1} \, \pi_j} f_j.
	\end{equation*}
	This, of course, implies that $\candidate{e_i}{f_j}$ holds; a contradiction is created because $(\finv, \binv)$ were assumed to satisfy Condition~\ref{con:separator}~\ref{sep2}.

	Therefore, to conclude the second direction of the proof, we have shown that the existence of a non-reachability witness $(I, J)$ for $\source$ and $\target$ indeed implies that $\source$ cannot reach $\target$ in $\Aa$.
\end{proof}

\complexity*
\begin{proof}
	We will use~\cref{lem:leaky-invariant} to prove that non-reachability in OCA with disequality tests is in $\sigmatwop$.
	For an instance of (non-)reachability, let $\Aa$ be an OCA with disequality tests, let $\source$ be the initial configuration, and let $\target$ be the target configuration.
	Suppose that there is no run from $\source$ to $\target$ in $\Aa$.
	By~\cref{lem:leaky-invariant}, there exists a non-reachability witness $(I, J)$ for $\source$ to $\target$ such that $I$ and $J$ have concise descriptions.
	Let $\finv \coloneqq \pposts{I} \cup \Post{\pposts{I}}$ and $\binv \coloneqq \opres{J} \cup \Pre{\opres{J}}$.
	We now argue that checking that $(I, J)$ satisfies the conditions to be a non-reachability witness can be completed in non-determinsitic polynomial time.
	\begin{enumerate}
		\item 
			In polynomial time, one can check whether $\source \in I$, $\target \in J$, $I \sset \posconfigs \cap L$, and $J \sset \rev{\posconfigs} \cap \rev{L}$.
			Recall that $I$ consists of at most $3\size{Q}+1$ arithmetic progressions; suppose $I = I_1 \cup \ldots \cup I_k$ where $k \leq 3\size{Q}+1$ and, for each $i \in [1, k]$, $I_i$ is an arithmetic progression on some state $q$.
			To check $I \sset \posconfigs \cap L$, we can check that $I_i \sset \posconfigs$ and $I_i \sset L$ for every $i \in [1, k]$.
			For convenience, suppose $(q, u)$ and $(q, v)$ are the configurations with the smallest and largest counter values in $I_i$ and suppose $d$ is the difference of $I_i$.
			First, to check that $I_i \sset \posconfigs$, it suffices to check that $(q, u) \in \posconfigs$; this can be done in polynomial time via~\cref{lem:define-canonical}.
			Second, to check that $I_i \sset L$, we will construct an instance of unboundedness in a strongly connected 1-VASS with disequality tests.
			Let $\Aa_S$ be the strongly connected component of $\Aa$ that $q$ belongs to.
			Add, to $\Aa_S$, an additional state $q'$ with a disequality test $x \neq v + d$ and two transitions $(q', 0, q)$ and $(q', d, q')$, to obtain a new OCA with disequality tests $\Bb$.
			It follows that $(q', u)$ is bounded in $\Bb$ if and only if $I_i \sset L$.
			This is true because if $(q', u)$ is unbounded in $\Bb$, then there is some $(q, x) \in I_i$ that is unbounded in $\Aa_S$ which means that $(q, x)$ is not locally bounded.
			Recall that (un)boundedness can be decided in polynomial time (see~\cref{decide_boundedness}).

		\item
			In non-deterministic polynomial time, one can check whether or not $(I, J)$ are inductive (\cref{con:inductive}).
			Verifying that $\Post{\bpposts{I}} \cap \posconfigs \cap L \not\sset I$ requires checking membership in $\posconfigs$ and $L$, which can be completed in polynomial time by~\cref{lem:define-canonical} and~\cref{decide_boundedness}, respectively.
			The same can be said for verifying \linebreak$\Pre{\bopres{J}} \cap \rev{\posconfigs} \cap \rev{L} \sset J$.

		\item
			In non-deterministic polynomial time, one can check whether or not $(\finv, \binv)$ satisfies the first condition of being a separator (\cref{con:separator}~\ref{sep1}).
			First, guess four configurations $c$, $c'$, $d'$, and $d$.
			One can check whether $c \in I$, $d \in J$, whether $c' \in \pposts{c}$ or $c' \in \Post{\pposts{c}}$ and whether $d' \in \opres{d}$ or $d' \in \Pre{\opres{d}}$ in non-determinsitic polynomial time (using~\cref{certificate_pessimistic_runs_cc}).
			In which case, in linear time, one can see whether~\ref{sep1} is violated by checking $c' \rightarrow d'$.

		\item
			In non-deterministic polynomial time, one can check whether or not $(\finv, \binv)$ satisfies the second condition of being a separator (\cref{con:separator}~\ref{sep2}).
			Similar to checking the first condition of being a separator, guess for configurations $c$, $c'$, $d'$, and $d$.
			Like for condition 3, one can check whether $c \in I$, $d \in J$, whether $c' \in \pposts{c}$ or $c' \in \Post{\pposts{c}}$ and whether $d' \in \opres{d}$ or $d' \in \Pre{\opres{d}}$ in non-determinsitic polynomial time (using~\cref{certificate_pessimistic_runs_cc}).
			However, in this case, one can check whether~\ref{sep2} is violated by checking if $c' \notin L$ and $d' \notin \rev{L}$ in polynomial time using~\cref{decide_boundedness} as well as checking $\candidate{c'}{d'}$ in non-deterministic polynomial time (say by using integer linear programming).
	\end{enumerate}
	Note that checking conditions 3 and 4 requires exponential bounds on the counter values of configurations in $\finv$ and $\binv$ (see~\cref{pessimistic_run_bounds}).
\end{proof}